\documentclass[useAMS,usenatbib]{mn2e}

 \usepackage{Times}

\usepackage{psfrag}
\usepackage{pspicture}
\usepackage{graphicx}

\newcommand\lsim{~\lower.5ex\hbox{$\buildrel < \over \sim$}~}
\newcommand\gsim{~\lower.5ex\hbox{$\buildrel > \over \sim$}~}

\def\gsim{~\lower.6ex\hbox{$\buildrel > \over \sim$}~}

\title[Star formation at $z=1.47$: a stereoscopic view]{Star formation at $\bf z=1.47$ from HiZELS: An H$\alpha$+[O{\sc ii}] double-blind study\thanks{Based on observations obtained using both the Wide Field CAmera on the 3.8m United Kingdom Infrared Telescope (UKIRT), as part of the High-$z$ Emission Line Survey (HiZELS), and Suprime-Cam on the 8.2-m Subaru Telescope, which is operated by the National Astronomical Observatory of Japan.} }
\author[D. Sobral et al.]{David Sobral$^{1,2}$\thanks{E-mail: sobral@strw.leidenuniv.nl}, Philip N. Best$^{1}$, Yuichi Matsuda$^{3}$, Ian Smail$^{3}$, James E. Geach$^{3,4}$
\newauthor and Michele Cirasuolo$^{1,5}$ \\
$^{1}$SUPA, Institute for Astronomy, Royal Observatory of Edinburgh, Blackford Hill, Edinburgh, EH9 3HJ, UK\\
$^{2}$Leiden Observatory, Leiden University, P.O.\ Box 9513, NL-2300 RA Leiden, The Netherlands\\
$^{3}$Institute for Computational Cosmology, Durham University, South Road, Durham, DH1 3LE, UK\\
$^{4}$Department of Physics, McGill University, Ernest Rutherford Building, 3600 Rue University, Montr\'eal, Qu\'ebec, Canada, H3A 2T8\\
$^{5}$Astronomy Technology Centre, Royal Observatory of Edinburgh, Blackford Hill, Edinburgh, EH9 3HJ, UK\\
}
\begin{document}
\date{Accepted 2011 October 9. Received 2011 October 4; in original form 2011 September 8}
\pagerange{\pageref{firstpage}--\pageref{lastpage}} \pubyear{2011}
\maketitle

\label{firstpage}
\begin{abstract}

This paper presents the results from the first wide and deep dual narrow-band survey to select H$\alpha$ and [O{\sc ii}] line emitters at $z=1.47\pm0.02$, exploiting synergies between the United Kingdom InfraRed Telescope and the Subaru telescope by using matched narrow-band filters in the $H$ and $z'$ bands. The H$\alpha$ survey at $z=1.47$ reaches a 3\,$\sigma$ flux limit of F$_{{\rm H}\alpha}\approx7\times10^{-17}$\,erg\,s$^{-1}$\,cm$^{-2}$ (corresponding to a limiting SFR in H$\alpha$ of $\approx7$\,M$_{\odot}$\,yr$^{-1}$) and detects $\approx200$ H$\alpha$ emitters over $0.7$\,deg$^2$, while the much deeper [O{\sc ii}] survey reaches an effective flux of $\approx7\times10^{-18}$\,erg\,s$^{-1}$\,cm$^{-2}$ (SFR in [O{\sc ii}] of $\sim1$ M$_{\odot}$\,yr$^{-1}$), detecting $\approx1400$ $z=1.47$ [O{\sc ii}] emitters in a matched co-moving volume of $\sim2.5 \times 10^5$\,Mpc$^3$. The combined survey results in the identification of 190 simultaneous H$\alpha$ and [O{\sc ii}] emitters at $z=1.47$. H$\alpha$ and [O{\sc ii}] luminosity functions are derived and both are shown to evolve significantly from $z\sim0$ in a consistent way. The star formation rate density of the Universe at $z=1.47$ is evaluated, with the H$\alpha$ analysis yielding $\rho_{\rm SFR}=0.16\pm0.05$\,M$_{\odot}$\,yr$^{-1}$ Mpc$^{-3}$ and the [O{\sc ii}] analysis $\rho_{\rm SFR}=0.17\pm0.04$\,M$_{\odot}$\,yr$^{-1}$ Mpc$^{-3}$. The measurements are combined with other studies, providing a self-consistent measurement of the star formation history of the Universe over the last $\sim11$\,Gyrs. By using a large comparison sample at $z\sim0.1$, derived from the Sloan Digital Sky Survey, [O{\sc ii}]/H$\alpha$ line ratios are calibrated as probes of dust-extinction. H$\alpha$ emitters at $z\sim1.47$ show on average A$_{\rm H\alpha}\approx1$\,mag, the same as found by SDSS in the local Universe. It is shown that although dust extinction correlates with SFR, the relation evolves by about $\sim0.5$\,mag from $z\sim1.5$ to $z\sim0$, with local relations over-predicting the dust extinction corrections at high-$z$ by that amount. Stellar mass is found to be a much more fundamental extinction predictor, with the same relation between mass and dust-extinction being valid at both $z\sim0$ and $z\sim1.5$, at least for low and moderate stellar masses. The evolution in the extinction-SFR relation is therefore interpreted as being due to the evolution in median specific SFRs over cosmic time. Dust extinction corrections as a function of optical colours are also derived and shown to be broadly valid at both $z\sim0$ and $z\sim1.5$, offering simpler mechanisms for estimating extinction in moderately star-forming systems over the last $\sim9$\,Gyrs.

\end{abstract}


\section{Introduction}\label{intro}

It is now clear that the ``epoch'' of galaxy formation occurs at
$z>1$, as surveys measuring the star formation rate density
($\rho_{\rm SFR}$) as a function of epoch show that $\rho_{\rm SFR}$
rises steeply out to $z\sim1$ \citep[e.g.][]{Lilly96,Hopkins2006}, but determining
the redshift where $\rho_{\rm SFR}$ peaked at $z>1$ is more difficult.

Accurately determining $\rho_{\rm SFR}$ requires the selection of
clean and well-defined samples of star-forming galaxies over representative volumes. In
practice, such selection is done by detecting signatures of massive stars
(being very short-lived, their presence implies recent episodes of
star formation). The high luminosities of such massive
stars allow them to be traced up to very high redshift, and to
estimate star formation rates (SFRs). Such stars emit
strongly in the ultra-violet (UV), and although this is often
significantly absorbed, it is then re-emitted through a variety of
processes, resulting in detectable signatures such as strong emission
lines or thermal infra-red emission from heated dust \citep[see
  discussion in][]{Hopkins_SLOAN}. Ideally, the use of different tracers
of recent star formation would provide a consistent view, but in
reality, because they have different selection biases, and require
different assumptions/extrapolations, measurements are often
significantly discrepant \citep[e.g.][]{Hopkins2006}. A major issue is
the difficulty and uncertainty in correcting for (dust) extinction,
especially for UV and optical wavelengths, leading to large systematic
uncertainties (and potentially missing entire populations). An
additional complication comes from surveys at different epochs making
use of different indicators due to instrumentation and detection
limitations. For example, while for $0<z \lsim 0.4$ the evolution of $\rho_{\rm
  SFR}$ is typically estimated using H$\alpha$ luminosity, because it
is easily targeted in the optical
\citep[e.g.][]{Ly2007,Shioya,Dale2010}, at higher redshifts the line is
redshifted into the near-infrared, and [O{\sc ii}]\,3727 -- hereafter
[O{\sc ii}] -- luminosity (much more affected by extinction than
H$\alpha$, and also a metallicity-dependent indicator) is used
instead \citep[e.g.][]{Zhu,Bayliss}.

The [O{\sc ii}] emission line can be traced in the optical window up
to $z\sim1.5$ and many authors have attempted to measure the evolution
of [O{\sc ii}] luminosity function up to such look-back times
\citep[e.g.][]{Teplitz03,Hogg,GalegoOII,Takahashi,Zhu}. Such studies have
identified a strong evolution from $z=0$ to $z>1$
\citep[e.g.][]{GalegoOII,Zhu}, although the evolution seems to be
slightly different from that seen in the H$\alpha$ luminosity
function. Part of the differences may well arise from the difficulty
in using [O{\sc ii}] luminosity density directly as a star-formation
rate density indicator: [O{\sc ii}] luminosity is calibrated locally using
H$\alpha$ \citep[see e.g.][for a comparison between both
  emission-lines at low-$z$]{ARAGON,Mouhcine}, but the actual
calibration depends on metallicity and dust-extinction
\citep[c.f.][]{Jansen,Kewley04}. These issues have been reported by
many studies, mostly using data from the local Universe or low
redshift \citep[see e.g.][for a comparison between both lines and
  other indicators in the Sloan Digital Sky Survey]{Gilbank10}, for which it is possible to
measure both H$\alpha$ and [O{\sc ii}]. However, so far it has not
been possible to directly compare the H$\alpha$ and [O{\sc ii}]
indicators at $z>1$ using large-enough samples to test, extend and
improve our understanding.

While some H$\alpha$ studies beyond $z\sim0.4$ have been conducted in the 90's \citep[e.g.][]{Bunker95,Malkan95}, tracing the H$\alpha$ emission line in the infrared, the small field-of-view of infrared detectors at that time made it very hard to detect more than $\sim1-2$ emitters at $z>1$ \citep[e.g.][]{PaulVan} using ground-based narrow-band surveys. Slitless spectroscopy using NICMOS on the Hubble Space Telescope (HST) provided a space-based alternative to make significant progress, particularly by allowing much deeper observations than those from the ground \citep[e.g.][]{McCarthy,Yan,Hopkins2000}. Further progress is now being obtained with WFC3 on HST, as recent studies \citep[e.g.][]{Atek2010,HSTTHER,vanDokkum} take advantage of the increased sensibility and the wider field-of-view (although still relatively small when compared to wide-field ground-based infrared detectors) of WFC3 to simultaneously look for fainter emitters and increase the sample sizes. Furthermore, the development of wide-field near-infrared cameras in the last decade has finally made it possible to conduct ground-based large area H$\alpha$ surveys (necessary to overcome cosmic variance) which can look for such emission-line galaxies all the way up to $z\approx2.5$.

HiZELS, the High-redshift(Z) Emission Line Survey\footnote{For more
  details on the survey, progress and data releases, see
  http://www.roe.ac.uk/ifa/HiZELS/} \citep[][hereafter S09]{G08,S09a} is a
Campaign Project using the Wide Field CAMera (WFCAM) on the United
Kingdom Infra-Red Telescope (UKIRT) and exploits specially-designed
narrow-band filters in the J and H bands (NB$_{\rm J}$ and NB$_{\rm H}$),
along with the H$_2$S1 filter in the K band, to undertake panoramic, moderate depth surveys for line emitters. HiZELS is primarily targeting
the H$\alpha$ emission line redshifted into the near-infrared at $z=0.84$, $z=1.47$
and $z=2.23$. The main HiZELS survey aims to cover $\approx7$\,deg$^2$ (to overcome cosmic variance)
and is detecting $\sim1000$ emitters over volumes of
$>10^6$\,Mpc$^3$ with each filter, reaching limiting H$\alpha$
(observed) SFRs of $\approx3$, $6$ and $30$\,M$_{\odot}$\,yr$^{-1}$
at $z=0.84$, $z=1.47$ and $z=2.23$. Such data will pin down
the likely peak of $\rho_{\rm SFR}$ and provide detailed information
about the population of star-forming galaxies at each epoch \citep[see][]{Best2010}.

HiZELS provides precise measurements of the evolution of the H$\alpha$
luminosity function from $z=0.0$ to $z=2.23$ (Geach et al. 2008;
S09), while the contribution from much deeper H$\alpha$ surveys over smaller areas \citep[e.g.][]{Hayes,CHU11} offers the possibility of also exploring the faint-population. The H$\alpha$ luminosity function evolves significantly, mostly
due to an increase by more than one order of magnitude in L$_{\rm
  H\alpha}^*$ (S09), the characteristic H$\alpha$ luminosity, from the local
Universe to $z=2.23$. In addition, \cite{SOBRAL10B} found that at
$z=0.84$ the faint-end slope of the luminosity function is strongly
dependent on the environment, with the H$\alpha$ luminosity function
being much steeper in low density regions and much shallower in the
group/cluster environments. Whether this is also found at higher
redshifts remains unknown.

%
%
\begin{figure}
\centering
\includegraphics[width=8.2cm]{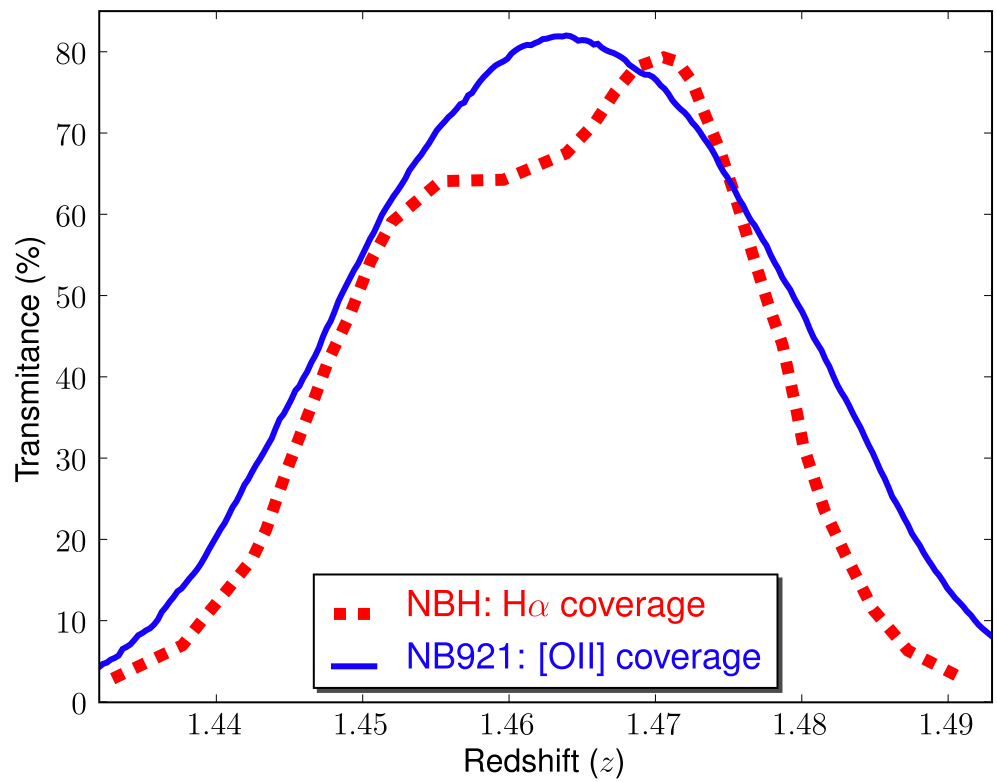}
\caption[A comparison between narrow-band filter profiles]{A
  comparison between the NB$_{\rm H}$ redshift coverage of the
  H$\alpha$ emission line and the NB921 redshift coverage of the
  [O{\sc ii}] emission-line, revealing that the two filters are
  extremely well matched and ideal for a dual H$\alpha$-[O{\sc ii}]
  narrow-band survey. The [O{\sc ii}] coverage is only slightly wider
  (in redshift) than the H$\alpha$ coverage -- but this is shown to
  have little to no effect on the analysis (c.f. Section
  \ref{filter_profiles}). \label{F_profiles}}
\end{figure}

Indeed, while it is now possible to detect H$\alpha$ emission over
very wide areas up to $z>2$, distinguishing between H$\alpha$ and
any other emission line at any redshift is a challenge, particularly
over large areas where it is unfeasible to assemble ultra-deep
multi-wavelength data over dozens of bands. Matched dual narrow-band
surveys offer a solution to the problem; in particular,
since the NB921 narrow-band filter on Subaru is able to probe the
[O{\sc ii}] emission line for the same redshift range ($z\sim1.47$) as
the HiZELS narrow-band $H$ filter on UKIRT probes H$\alpha$ (see Figure
\ref{F_profiles}), a matched and sufficiently deep survey would not
only provide a simple, clean selection ($z=1.47$ sources will be
detected as emitters in both data-sets), but also provide a means of
directly comparing H$\alpha$ and [O{\sc ii}] at $z\sim1.5$ for large samples for the first time.

%
%
\begin{table*}
 \centering
  \caption{Observation log for the NB$_{\rm H}$ observations of the
    UDS field, taken using WFCAM on UKIRT, during 2008, 2009 and
    2010. Limiting magnitudes (3\,$\sigma$) are the average of the
    four frames which constitute a field, based on the the
    measurements of 10$^6$ randomly placed 2$''$ apertures in each
    frame. Integration times given in brackets present the total
    integration times obtained at the telescope prior to 
    rejection of data taken in poorer conditions (see \S\ref{observationes}).}
  \begin{tabular}{@{}ccccccccc@{}}
  \hline
   Field & R.A. & Dec. & Int.\ time & FHWM  & Dates  & $m_{lim}$ NB$_{\rm H (Vega)}$  \\
        & {(J2000)} &(J2000) & (ks) & ($''$) &  & (3$\sigma$) \\
 \hline
   \noalign{\smallskip}
UKIDSS-UDS NE & 02\,18\,29 & $-$04\,52\,20 & 18.2 (18.2) & 0.8 & 2008 Sep 28-29; 2009 Aug 16-17; 2010 Jul 22 & 21.2 \\
UKIDSS-UDS NW & 02\,17\,36 & $-$04\,52\,20 & 17.1 (18.3) & 0.9 & 2008 Sep 25, 29; 2010 Jul 18, 22 & 20.9\\
UKIDSS-UDS SE & 02\,18\,29 & $-$05\,05\,53 & 28.0 (28.0) & 0.8 & 2008 Sep 25, 28-29; 2009 Aug 16-17 & 21.4  \\
UKIDSS-UDS SW  &  02\,17\,38 & $-$05\,05\,34 & 18.3 (19.1) & 0.8 & 2008 Oct-Nov; 2009 Aug 16-17; 2010 Jul 23 & 21.2  \\
 \hline
\end{tabular}
\label{obs}
\end{table*}

This paper presents deep narrow-band imaging using the NB$_{\rm H}$
filter at $\lambda = 1.617\umu$m, as part of HiZELS, over $0.79\deg^2$
in the UKIDSS Ultra Deep Survey \citep{Lawrence}
field (UDS) and combines the data with ultra-deep NB921 imaging taken
using Suprime-Cam on the Subaru telescope, to explore an extremely well-matched H$\alpha$-[O{\sc ii}] narrow-band survey over $0.70\deg^2$.

The paper is organised as follows: \S2 presents the
observations, data reduction, source extraction and catalogue
production. In \S3, emission-line galaxies are selected and, using
colours and photometric redshifts, the samples of H$\alpha$ and [O{\sc
    ii}] selected $z=1.47$ emitters are presented. \S4 presents the
H$\alpha$ and [O{\sc ii}] luminosity functions and the derived 
star-formation rate density at $z=1.47$,
together with an accurate measurement of their evolution. In \S5, the
large sample of robust H$\alpha$-[O{\sc ii}] emitters is used to
conduct line ratio studies and compare these with a large sample selected
locally from the Sloan Digital Sky Survey (SDSS). These are used to
investigate the extinction properties of the galaxies, and how these
evolve with redshift.  Finally, \S6
outlines the main conclusions. An H$_0=70$\,km\,s$^{-1}$\,Mpc$^{-1}$,
$\Omega_M=0.3$ and $\Omega_{\Lambda}=0.7$ cosmology is
used. Magnitudes are in the Vega system, except if noted otherwise.

\section{DATA AND SAMPLES}\label{data_technique}

\subsection{Near-infrared NB$_{\rm H}$ imaging with UKIRT}\label{observationes}

The UKIDSS UDS field was observed with WFCAM on UKIRT using a set of
custom narrow-band $H$ filters (NB$_{\rm H}$, $\lambda =
1.617\,\umu$m, $\Delta\lambda = 0.021\,\umu$m), as detailed in Table 1. WFCAM's standard ``paw-print'' configuration of four
$2048\times2048$ ($0.4''$\,pixel$^{-1}$) detectors offset by $\sim20'$
was macrostepped four times to cover a contiguous region of
$\sim55'\times55'$ \citep{Casali}, with individual narrow-band
exposures of 100\,s. The conditions were mostly photometric. The Non
Destructive Read (NDR) mode was used for all narrow-band observations
to minimise the effects of cosmic rays in long exposures. The
observations were obtained by jittering around 14 different positions
in each of the 4 pointings. The UKIDSS UDS $H$-band image (24\,AB, 5$\sigma$) overlaps
entirely with the full narrow-band image, yielding a total
survey area of $0.79 \deg^2$.

A dedicated pipeline has been developed for HiZELS (PfHiZELS, c.f. S09
for more details), and was used to fully reduce the data. Very
briefly, the pipeline dark-subtracts the data and median combines
dark-subtracted data (without offsetting) to obtain a first-pass flat
field for each jitter pattern. The latter is used to produce a
badpixel mask for each chip by flagging pixels which deviate by more
than $5\sigma$ from the median value. Frames are individually
flattened and individual source masks are produced for each frame. A
new final flat is then created masking detected sources (for each
jitter pattern) and all data are flattened. A world coordinate system
is fitted to each frame by querying the USNO A2.0 catalogue (typically
$\sim70-80$ stars per frame). Finally, all individual reduced frames
were visually inspected, resulting in the rejection (for each of the
four cameras) of 12 frames for the NW field and 8 frames in the SW
field due to bad quality (FWHM$>1.1''$ or $A_H>0.25$ mag). Final
frames which passed the data-control were de-jittered and median
co-added with {\sc SWarp} (Bertin 1998), including a background
mesh-based sky subtraction optimized for narrow-band data, and masking
of bad-pixels and cosmic-rays. It should be noted that WFCAM frames
suffer from significant cross-talk artifacts (toroidal features at
regular pixel intervals from bright sources in the read-out
direction); because these are linked to a ``physical'' location, they
can only be removed from the source catalogue (see \S\ref{extraction}).

Narrow-band images were photometrically calibrated (independently) by
matching $\sim70$ stars with $m_H=11-16$ per frame from the 2MASS
All-Sky catalogue of Point Sources \citep{2MASS} which are unsaturated
in the narrow-band frames.

\subsection{NB$_{\rm H}$ Source Extraction and Survey Limits}\label{extraction}

Sources were extracted using {\sc SExtractor}
\citep{SExtractor}. Photometry was measured in apertures of $2''$
diameter. In order to clean spurious sources from the catalogue
(essential to remove cross-talk artifacts), the final images were
visually inspected; this revealed that sources brighter than
$\sim11.5$\,mag (NB$_{\rm H}$, Vega) were surrounded by a number of artifacts
detected within ``bright halos'', as well as cross-talk. Fainter
sources (up to 16\,mag) showed only cross-talk features. Sources
fainter than 16\,mag did not produce any detectable cross-talk at the
depth of the observations. Cross-talk sources and detections in the
halo regions were removed from the catalogue separately for each
frame, which greatly simplifies their identification.

The average 3\,$\sigma$ depth of the entire set of NB$_{\rm H}$ frames
is 21.2\,mag; this is measured using a set of 10$^6$ randomly placed
apertures per frame. Above the 3\,$\sigma$ threshold in each frame,
the narrow-band imaging detects a total of 23394 sources (5904, 4533,
6946 and 6011 in the NE, NW, SE and SW pointings, respectively) across $0.78 \deg^2$ (after removal of regions in which cross-talk and
other artifacts caused by bright objects are located)

%
%
\begin{figure*}
\begin{minipage}[b]{0.48\linewidth}
\centering
\includegraphics[width=8.2cm]{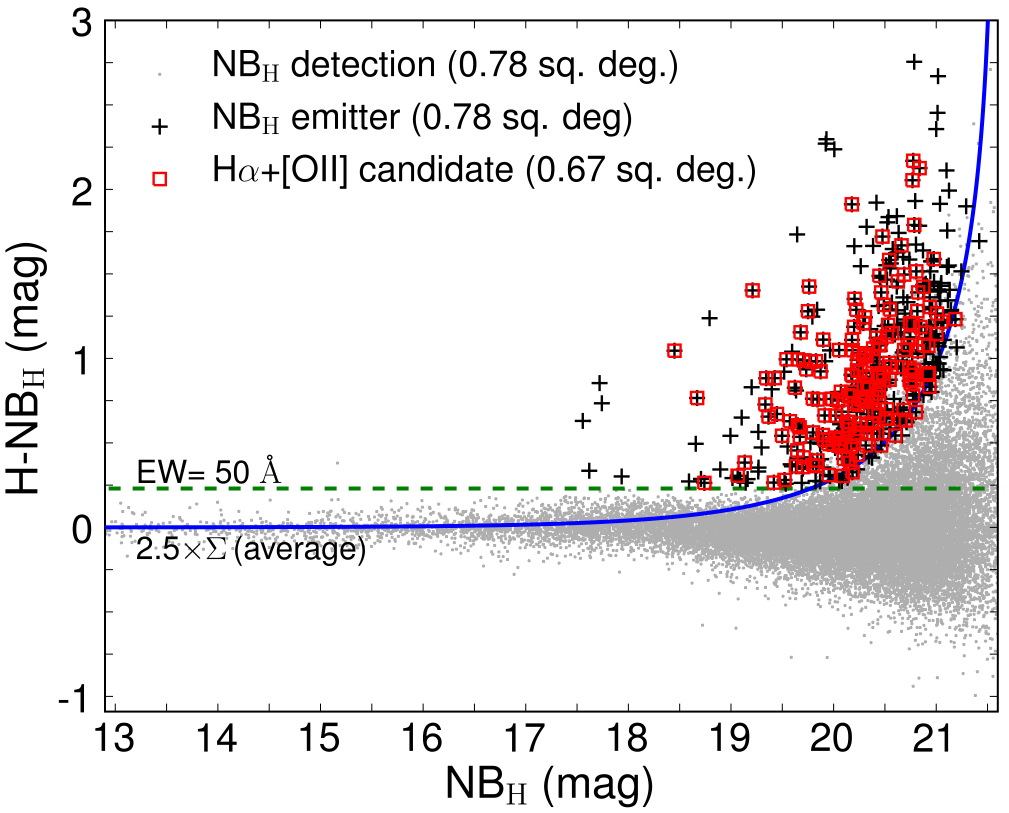}
\end{minipage}
\hspace{0.1cm}
\begin{minipage}[b]{0.48\linewidth}
\centering
\includegraphics[width=8.2cm]{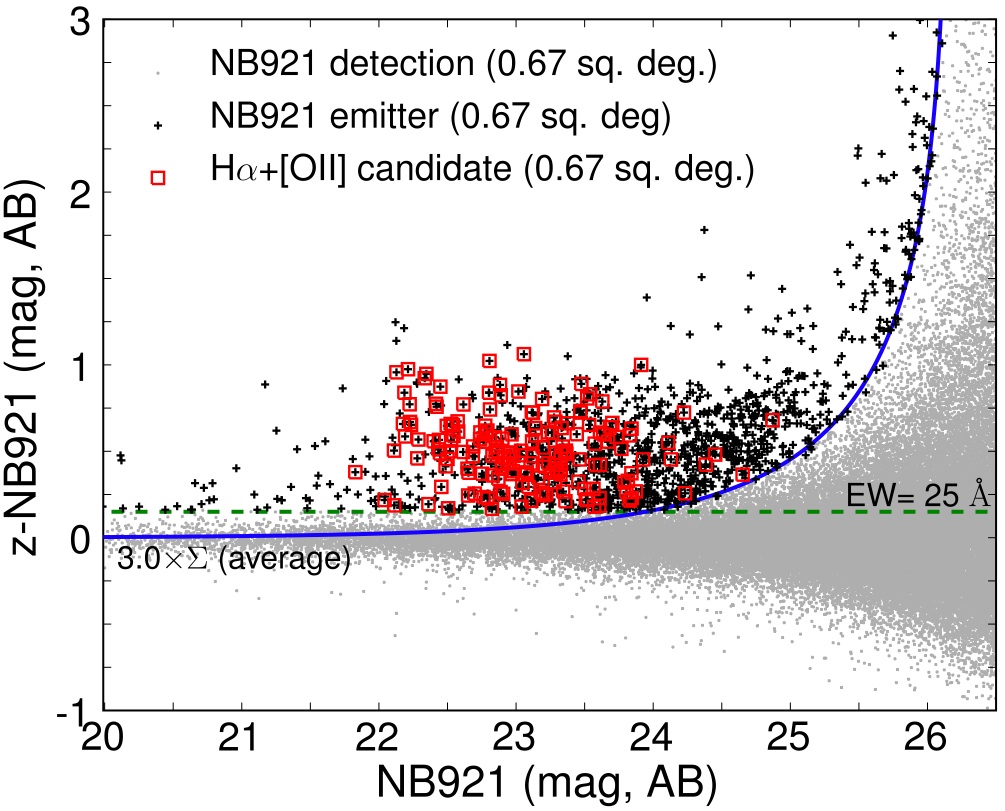}
\end{minipage}
\caption[Selection of potential emitters with UKIRT and Subaru]{Narrow-band excess as a function of narrow-band magnitude for NB$_{\rm H}$ (left panel, Vega magnitudes) and NB921 (right panel, AB magnitudes) data. These show $>3\sigma$ detections in narrow-band imaging and the lines present the average 2.5\,$\Sigma$ and 3.0\,$\Sigma$ colour significances for NB$_{\rm H}$ and NB921, respectively (for the average depth, but note that the analysis is done individually for each frame). Also, due to the very high number of detected sources in NB921, only 1 in every 5 sources is shown. The horizontal dashed lines present the equivalent width cuts used for NB$_{\rm H}$ and NB921 data -- these correspond to $z=1.47$ rest-frame EW limits of 20\,\AA \ for H$\alpha$ and 10\,\AA \ for [O{\sc ii}]. Narrow-band excess sources in both NB$_{\rm H}$ and NB921 (double-emitters) are shown, and constitute a robust sample of H$\alpha$/[O{\sc ii}] candidate emitters -- see Figure \ref{excess} for a visualisation of one such emitters. \label{colour-magC}}
\end{figure*}

\subsection{Optical NB921 imaging with Subaru}\label{NB921_imaging}

Archival Subaru/Suprime-Cam NB921 data of the UDS field are available,
taken by \cite{Ouchi2009, Ouchi2010}. The field was observed with
Suprime-Cam on Subaru as part of the Subaru/XMM-Newton Deep Survey
\citep[SXDS,][]{Ouchi2008} during 2005--2007. The NB921 filter is
centered at 9196\,\AA \ with a FWHM of 132\,\AA, and was used to cover
the field with 5 pointings with a total integration of 45.1 hours
(individual exposures ranging from 8 to 10 hours) -- see
\cite{Ouchi2010}. The raw NB921 data were downloaded from the archive
and reduced with the Suprime-Cam Deep field REDuction package
\citep[SDFRED,][]{Yagi2002,Ouchi2004} and IRAF. The combined images
were aligned to the public $z'$-band images of Subaru-XMM Deep Survey
\citep[SXDS;][]{Furusawa} and PSF matched (FWHM$=0.8''$). The NB921
zero points were determined using $z'$ data, so that the $z'$-NB921 colour
distributions of SXDS would be consistent with that of Subaru Deep
Field data \citep[][]{Kashikawa}. Source detection and photometry
were performed using {\sc SExtractor} \citep{SExtractor}. Sources were
detected on each individual NB921 image and magnitudes measured with
$2''$ diameter apertures. The average NB921 3$\sigma$ limiting
magnitude is estimated to be 26.3 (AB) by randomly placing $10^6$
2$''$ apertures in each frame; down to that depth, 347341 sources are
detected over the entire 1\,deg$^2$ area (5 Suprime-Cam pointings). Note that NB921 and $z'$ magnitudes are given in the AB system.

\section{SELECTION}\label{SELECTION}

\subsection{Narrowband excess selection}\label{narrowB_exc_selection}

Potential line emitters (NB$_{\rm H}$ and NB921) are selected
according to the significance of their broad-band $-$ narrow-band
($BB-NB$) colour, as they will have $(BB-\,NB)>0$. However, true
emitters need to be distinguished from those with positive colours due
to the scatter in the magnitude measurements; this is done by
quantifying the significance of the narrowband excess. The parameter
$\Sigma$ (see S09) quantifies the excess compared to the random scatter expected
for a source with zero colour, as a function of narrow-band magnitude
(c.f. S09).

%
%
\begin{figure}
\centering
\includegraphics[width=8.2cm]{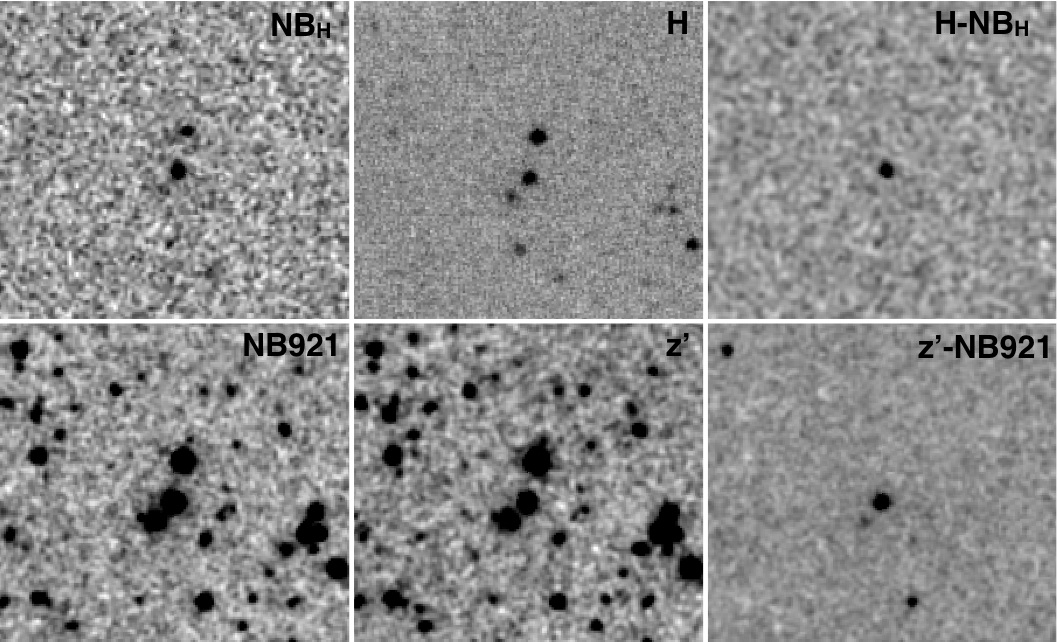}
\caption[Example of the double narrow-band technique]{The top panels
  show a strong NB$_{\rm H}$ emitter, clearly revealed after
  subtracting the $H$ continuum. The bottom panels show the same sky
  area ($\sim30''\times30''$) in NB921 and $z'$, revealing that the
  NB$_{\rm H}$ emitter is also a strong NB921 emitter (and thus a
  H$\alpha$+[O{\sc ii}] $z\sim1.47$ emitter). Note the depth
  difference, and how non-emitters disappear when the continuum
  estimated flux is subtracted from the narrow-band flux (other
  fainter NB921 emitters are also revealed). \label{excess}}
\end{figure}

Neither of the narrow-band filters falls at the center of the
broad-band filters and thus objects with redder colours will tend to
have a negative ($H-$\,NB$_{\rm H}$) colour, and a positive ($z'$-NB921) colour,
while bluer sources will have ($H-$\,NB$_{\rm H})>0$ and ($z'$-NB921$)<0$. This
will not only affect the selection of emission line objects, but can
also result in an over/under-estimation of emission-line fluxes
because it will lead to an under/over-estimation of the
continuum. Fortunately, this can be broadly corrected by considering
the broad-band colours of each source, as in S09. For the NB$_{\rm H}$
data, this is done by studying ($H-$\,NB$_{\rm H}$) as a function of
$(J-H)$ colour for all sources detected in the NB$_{\rm H}$ frames. 
 A linear fit is derived\footnote{Only sources within
  $\pm2\sigma$ of the general scatter around the median $H-$\,NB$_{\rm
    H}$ are used. Furthermore, in order to improve the fit for
  galaxies, potential stars (selected as sources satisfying
  $B-z>13.5(H-K)+2.0$, AB) are excluded and a separate fit is done just
  for the latter.} and is used to correct the initial $H$
magnitudes to produce an effective $H'$ magnitude appropriate for
estimation of the continuum contribution at the wavelength of the
NB$_{\rm H}$ filter. This assures a mean zero ($H'-$\,NB$_{\rm H}$) as a function of
($J-H$), and also results in no significant trend as a function of
($H-K$). The correction is given by:
\begin{equation}
(H'-\,\rm NB_H)=(\it H-\rm NB_H)-0.1(\it J-H\rm)-0.03.
\end{equation}
For the NB921 data, a similar approach is taken, but using $z'-J$
colours instead\footnote{For sources with no clear $J$ detection, a
  statistical correction of +0.05 is applied.}. Colours are corrected
by using the best linear fit:
\begin{equation}
(z''-\,\rm NB921)=(\it z'\rm-NB921)-0.05(\it z'-J'\rm)+0.15. 
\end{equation}
Emission line fluxes, F$_{{\rm line}}$, and equivalent widths,
EW$_{{\rm line}}$, are computed using:
\begin{equation}
   {\rm F}_{{\rm line}}=\Delta\lambda_{NB}\frac{f_{NB}-f_{BB}}{1-(\Delta\lambda_{NB}/\Delta\lambda_{BB})}
\end{equation}
and
\begin{equation}
   {\rm EW}_{{\rm line}}=\Delta\lambda_{NB}\frac{f_{NB}-f_{BB}}{f_{BB}-f_{NB}(\Delta\lambda_{NB}/\Delta\lambda_{BB})}
\end{equation}
where $\Delta\lambda_{NB}$ and $\Delta\lambda_{BB}$ are the FWHMs of
the narrow- and broad-band filters ($211$\,\AA \ and $132$\,\AA \ for
NB$_{\rm H}$ and NB921; $2893$\,\AA \ and $955$\,\AA \ for $H$ and
$z'$, respectively), and $f_{NB}$ and $f_{BB}$ are the flux densities
measured for the narrow and broad-bands, respectively. Note that $f_{BB}$ is computed by using the
corrected BB$'$ magnitudes, which are a much better approximation of the
continuum for the present purposes and guarantee that the median flux
distribution is zero. Note that the latter formula is only valid
because magnitudes and colours have been corrected to guarantee a flux
and colour distribution centered at zero.

The selection of emission-line candidates is done following S09
\citep[see also][]{Ouchi2010}. Narrow-band sources in the $H$ band
(NB$_{\rm H}$) are selected as line emitters if they have an individual $>3$\,$\sigma$ detection in NB$_{\rm H}$ and present a colour excess significance of $\Sigma>2.5$ (which broadly corresponds to a
flux limit, see S09), and an observed $\rm EW>50$\,\AA \ (corresponding to $\rm
EW>20$\,\AA \ rest-frame for the H$\alpha$ line at $z=1.47$). NB921
sources are selected as potential emitters if they have an individual $>3$\,$\sigma$ detection in NB921 and present a colour
excess significance of $\Sigma>3.0$ (as the data are much deeper, a
higher significance cut can be applied) and EW$>25$\,\AA \ (corresponding
to $\rm EW>10$\,\AA \ for [O{\sc ii}] emitters at $z=1.47$). The EW
cuts are applied to avoid including bright foreground objects with a
large significance and a steep continuum across the $H$ or $z'$ bands,
and were chosen to reflect the general scatter around the zero colour
at bright magnitudes (thus the difference between NB$_{\rm H}$ and
NB921) and to allow a good selection of both H$\alpha$ and [O{\sc ii}]
emitters. Figure \ref{colour-magC} presents the corrected broad-band
$-$ narrow-band colours as a function of narrow-band magnitude,
including the selection criteria and the sample of NB$_{\rm H}$ and
NB921 emitters. Figure \ref{excess} presents examples of emitters
drawn from the samples (for each filter).

%
%
\begin{figure*}
\begin{minipage}[b]{0.48\linewidth}
\centering
\includegraphics[width=8.2cm]{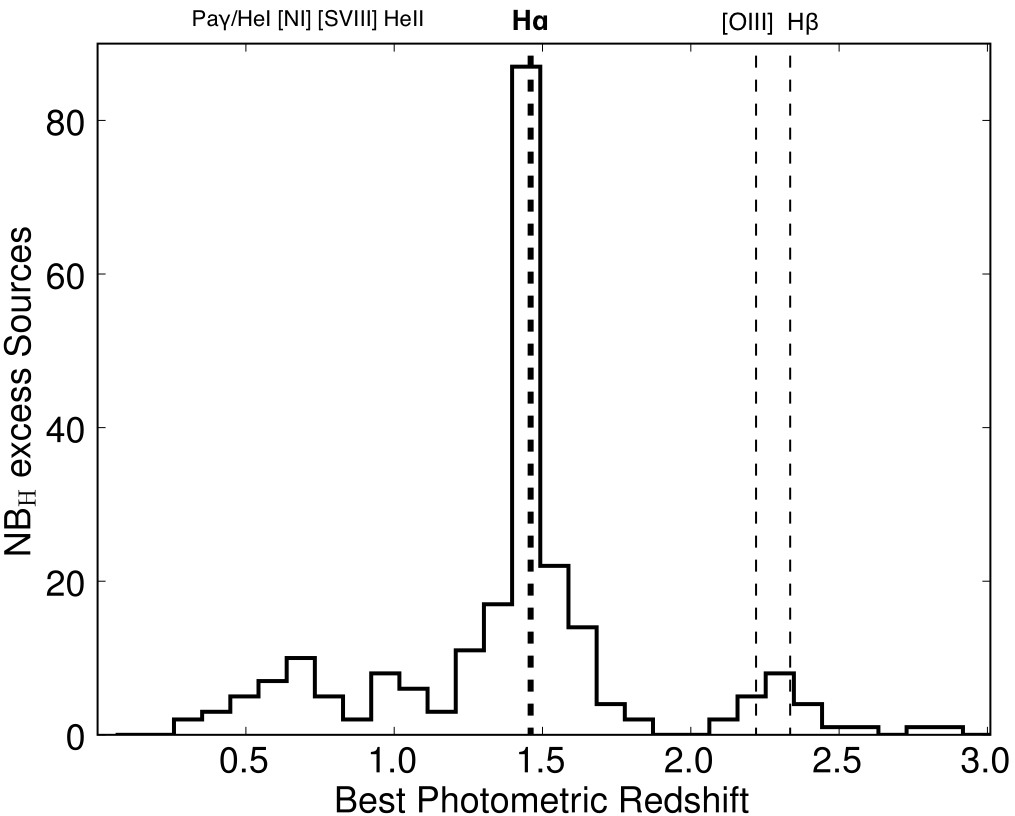}
\end{minipage}
\hspace{0.1cm}
\begin{minipage}[b]{0.48\linewidth}
\centering
\includegraphics[width=8.32cm]{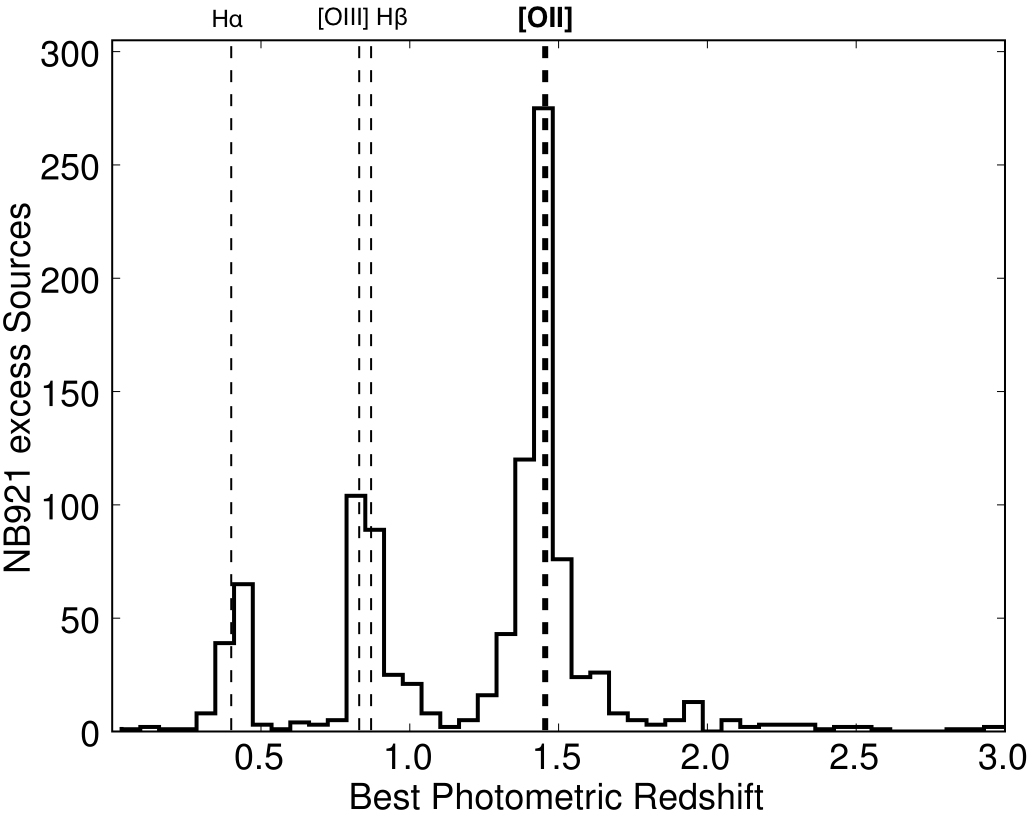}
\end{minipage}
\caption[The distirbution of photometric redshifts of both
  samples]{$Left$: photometric redshift distribution (peak of the
  probability distribution function for each source) for the NB$_{\rm
    H}$ emitter candidates. $Right$: the equivalent distribution for
  NB921 emitter candidates. Both distributions peak at $z\sim1.47$,
  corresponding to H$\alpha$ and [O{\sc ii}] emitters, respectively,
  but other populations of emitters are also found, such as H$\alpha$
  at $z=0.4$ and H$\beta$/[O{\sc iii}] at $z\sim0.83$ for the NB921
  data and Paschen-lines at $z<1$, [O{\sc iii}] at $z\sim2.2$ and
  [O{\sc ii}] at $z\sim3$ for the NB$_{\rm H}$ data. Ly$\alpha$
  emitter candidates in the NB921 dataset (51 with reliable
  photometric redshift within the SpUDS coverage) with photo-z of
  $z\sim6-7$ are not shown. \label{photoz}}
\end{figure*}

\subsection{The samples of narrow-band emitters}\label{NB_emitters}

Narrow-band detections below the estimated 3\,$\sigma$ threshold were not
considered. Due to the depth of the UKIDSS-UDS data (DR5), all
extracted NB$_{\rm H}$ sources have $>3\,\sigma$ broad-band $H$
detections (down to $\approx24.9$\,AB); for the much deeper NB921
data, only sources with $>3\sigma$ detections in $z'$ ($z'<26.7$) are
considered\footnote{This results in rejecting 5 per cent of the total number of potential emitters, but it should be noted that visual inspection of a sub-sample of these sources show that the majority are likely to be spurious, due to the combination of a faint detection in NB921 and a $<3$\,$\sigma$ limit in $z$. Furthermore, the results in this paper remain unchanged even if these sources are included in the analysis.}. The average 3\,$\sigma$ line flux limit is
$7\times10^{-17}$\,erg\,s$^{-1}$\,cm$^{-2}$ for the NB$_{\rm H}$ data
and $7\times10^{-18}$\,erg\,s$^{-1}$\,cm$^{-2}$ for the NB921
data. The first-pass NB$_{\rm H}$ sample of potential emitters contains 439 excess candidates out of all 23394 NB$_{\rm H}$ detections in the entire NB$_{\rm H}$ area
(0.78\,deg$^2$), while the NB921 sample has 8865 potential emitters out of
347341 NB921 individual detections over the entire NB921 area coverage. Table \ref{numbers} provides a summary of the number of sources and emitters throughout the selection proccess.

\subsubsection{Visual inspection and star exclusion} \label{VIsual}

All NB$_{\rm H}$ potential emitters are visually inspected in both the
broad-band and narrow-band imaging. Twenty six (26) sources were
removed from the sample as they were flagged as likely spurious. The
majority of these (15) correspond to artifacts caused by bright stars
that are on the edges of two or more frames simultaneously. The
remaining 11 sources removed were low S/N detections in noisy regions
of the NB$_{\rm H}$ image. After this visual check, the sample of potential NB$_{\rm H}$ is reduced to 413.

Even by applying a conservative EW cut, the sample of potential
emitters can be contaminated by stars. Fortunately, stars can be easily identified by
using the high-quality, multi-wavelength colour information
available for the SXDF-UDS field. In particular, an optical colour
vs. near-infrared colour (e.g. $B-z$ vs. $H-K$) is able to easily
separate stars from galaxies. Here, sources satisfying
$(B-z)>13.5(H-K)+2.0$ (AB) are classed as stars. By doing this, 2 potential
stars are identified in the sample of potential emitters selected from
the NB$_{\rm H}$ data, and 118 (out of 5623 emitters) from the NB921
data (the higher number of stars in the NB921 sample is driven both
by a much larger sample size and, more importantly, because of the lower EW cut used
to probe down to weak [O{\sc ii}] EWs). These are excluded
from the following analysis. It should be noted that all these sources
present ($B-z$) colours significantly larger than $13.5(H-K)+2.0$ (AB), and
therefore the identification of these as potential stars is not
affected by small changes in the separation criteria. One of the
potential NB$_{\rm H}$ emitters flagged as a star is likely to be a
cool T-dwarf \citep[see][]{S09b}.

The final sample of NB$_{\rm H}$
emitters over the entire NB$_{\rm H}$ area contains 411 sources, of which 135, 69, 136 and 71 are found
over the NE, NW, SE and SW fields, respectively
($\approx530$\,$\deg^{-2}$ over the entire field down to the average
NB$_{\rm H}$ UDS depth). Note that by restricting the analysis to the UKIDSS-UDS area coverage (matched to
the NB$_{\rm H}$ and NB921 simultaneous coverage), which will be mostly used throughout this paper, the survey covers 0.67\,deg$^2$; the sample of NB$_{\rm H}$ emitters is reduced to 295 sources (the rest of the sources are outside this area), while the sample of NB921 emitters has 5505 sources.

\subsection{Distinguishing between different line emitters} \label{foto_analysis}

There are many possible emission-lines which can be detected
individually by the NB$_{\rm H}$ and the NB921 filters. For $z<1$
galaxies, the NB$_{\rm H}$ filter is sensitive to lines such as
Pa$\beta$ at $z=0.26$ or Pa$\gamma$ and HeI at $z=0.49$ or HeII
at $z=0.96$, while at $z>1$ the (main) possible emission lines are
H$\alpha$ at $z\sim1.47$, [O{\sc iii}]/H$\beta$ at $z\sim2.25$ and
[O{\sc ii}] at $z=3.34$, among others. The NB921 filter is mostly
sensitive to H$\alpha$ at $z\sim0.4$, [O{\sc iii}]/H$\beta$ at
$z\sim0.83$, [O{\sc ii}] at $z\sim1.47$ and Ly$\alpha$ at
$z\sim6.6$. H$\alpha$--[O{\sc ii}] line emitters at $z=1.47$ can be
selected with a significant narrow-band excess in both bands, and thus
simultaneous excess sources provide an extremely robust means of
selecting a $z=1.47$ sample. This is explained in \S \ref{doube_approach}. First,
in order to select [O{\sc ii}] emitters below the H$\alpha$ flux
limit (see Figure \ref{colour-magC}), to assess the robustness and contamination of different
selections, to investigate the range of different emitters, and to allow the selection in sky areas where coverage is not available in both filters, alternative approaches are considered.

%
%
\begin{figure}
\centering \includegraphics[width=8.2cm]{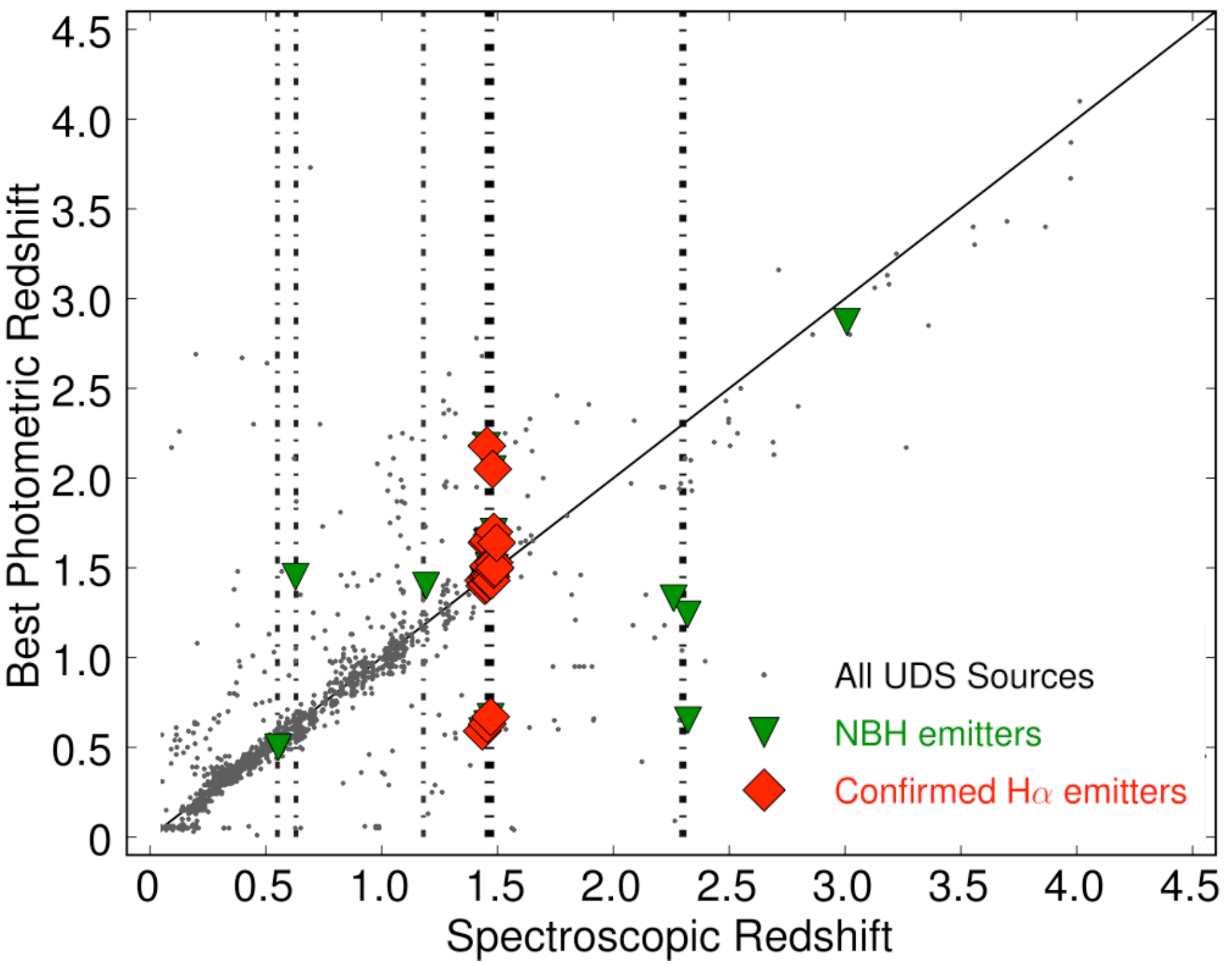}
\caption[Spectroscopic redshift distribution]{A comparison between the
  best photometric redshifts in the UDS field and the robust
  spectroscopic redshift for such sources in UDS with spectroscopic
  redshift. NB$_{\rm H}$ excess sources with spectroscopic redshifts
  are all identified with a known emission line falling into the
  NB$_{\rm H}$ filter. It is noteworthy that the limited (and very
  biased towards the strongest line emitters and AGN) spectroscopic
  sample reveals that some confirmed H$\alpha$ emitters have photo-zs
  which place them at a higher and lower redshift -- the use of NB921
  data is able to identify these emitters which the photo-zs could
  miss. Vertical lines indicate the redshifts of the main emission lines detected by the NB$_{\rm H}$ filter, specifically [N{\sc i}] at $z=0.545$, [S{\sc viii}]\,9914 at $z=0.625$, [O{\sc ii}]\,7625 at $z=1.125$, H$\alpha$ at $z=1.47$, [O{\sc iii}]\,5007 at $z=2.23$ and H$\beta$ at $z=2.33$. \label{spec_vs_photo}}
\end{figure}

\subsubsection{Photometric redshift analysis} \label{foto_analysis}

Multi-wavelength data can be used to effectively distinguish between
line-emitters at different redshifts (different emission lines) and
separately obtain H$\alpha$ and [O{\sc ii}] selected samples of
galaxies at $z=1.47$ from the NB$_{\rm H}$ and NB921 data-sets.  Out
of 5505 NB921 potential emitters (excluding stars) within the UKIDSS UDS area, 2715
($\approx50$\,\%) have a $>3$\,$\sigma$ detection in the near-IR
UKIDSS data ($K \lsim 25$\,AB), while all NB$_{\rm H}$ emitters are
detected in $K$. Robust photometric
redshifts\footnote{The photometric redshifts for UDS present
  $\sigma(\Delta z )=0.015$, where $\Delta z$ =
  $(z_{phot}-z_{spec})/(1+z_{spec})$. The fraction of outliers,
  defined as sources with $\Delta z>3\sigma(\Delta z)$, is lower than
  2\%. These photometric redshifts were used in
  \cite{SOBRAL10A,SOBRAL10B}.} are only available over a matched
$0.64\deg^2$ area, mostly due to the overlap with the Subaru data and
the deep Spitzer coverage (SpUDS, PI J. Dunlop), and for sources with
$K<23$ (M. Cirasuolo et al., in prep.). Photo-zs are derived using a range of photometry (\it UBVRizJHK \rm, together with IRAC 3.6 and 4.5 bands) and are available for 257\footnote{The area reduction by itself results in the loss
of 297 NB921 emitters and reduces the sample of NB$_{\rm H}$ line emitters
from 411 to 295. It should be
noted, nonetheless, that faint sources (with no reliable photo-z
information) and those outside the photo-z area can still be further
investigated by using colour-colour criteria.} NB$_{\rm H}$ potential line emitters and 1021
NB921 excess sources.

%
%
\begin{figure}
\centering
\includegraphics[width=8.2cm]{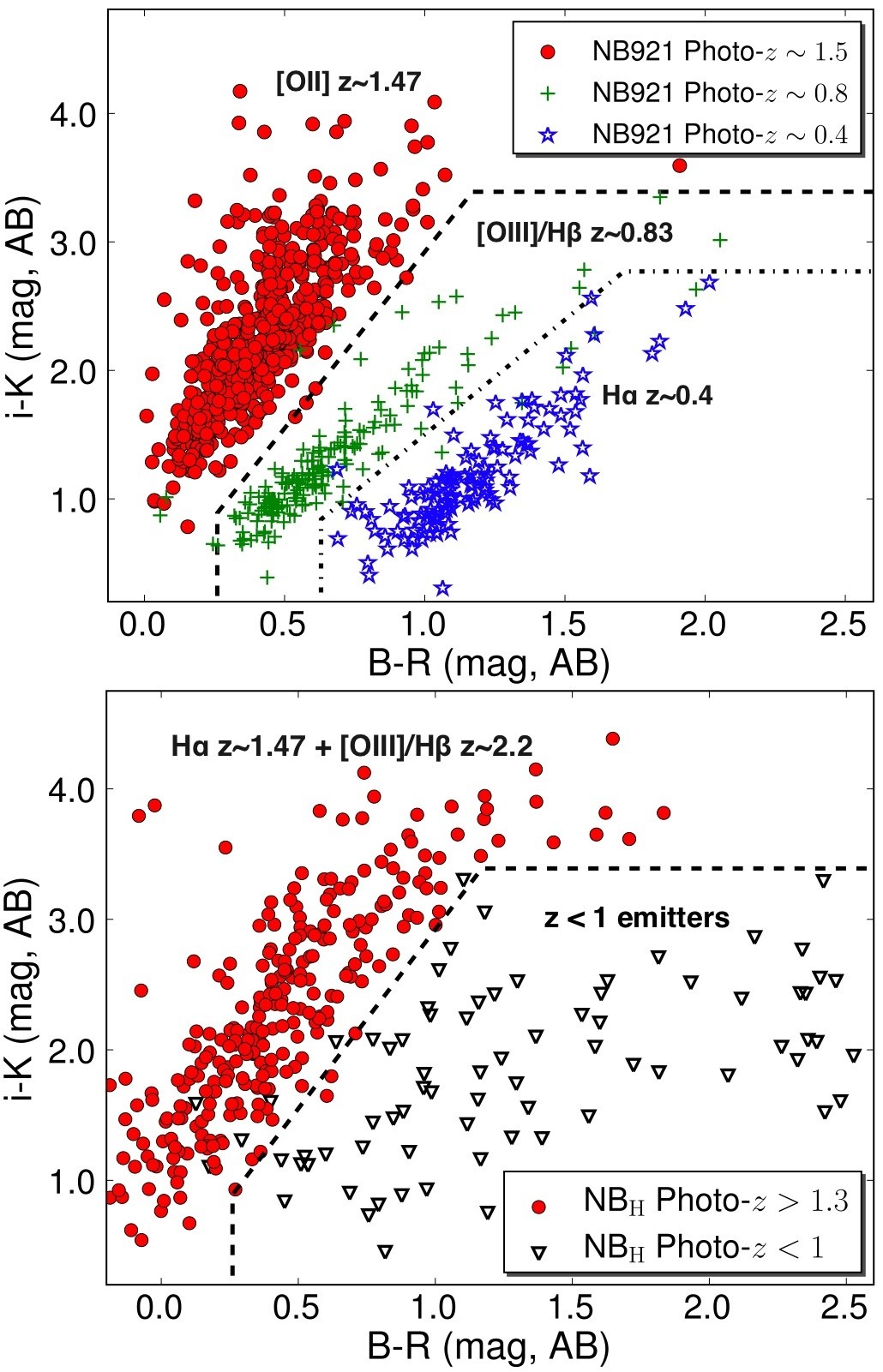}
\caption[The BRiK colour-colour separation]{$Top$: The $i-$K vs. B$-R$ (AB)
  colour-colour distribution of NB921 emitters, clearly separating the
  3 types of emitters that make the bulk of the sample of
  emitters. Simple lines for distinguishing these emitters --
  supported by evolution tracks (see S09) -- are also shown. $Bottom$:
  An equivalent plot for the NB$_{\rm H}$ emitters (AB), again
  demonstrating that lower redshift emitters can be easily isolated
  from the $z=1.47$ H$\alpha$ emitters. Higher redshift emitters,
  however, occupy a similar region, and can not be robustly isolated
  with this set of colours alone. The two pivot points for the $z\sim0.4$ and $z\sim0.8$ separation are [0.63,0.84] and [1.71,2.77]; while the two pivot points used to separate $z\sim0.8$ from $z\sim1.5$ are [0.26,0.89] and [1.17,3.39]. \label{colour_brik}}
\end{figure}

Figure~\ref{photoz} shows the photometric redshift distribution for
the selected narrow-band emitters in NB$_{\rm H}$ (left panel) and
NB921 (right panel) with available photometric redshifts in UDS, demonstrating the common peak at
$z\sim1.4-1.5$, associated with the H$\alpha$/[O{\sc ii}] lines being
detected in each narrow-band filter. In addition to this, the other
peaks are easily identified as H$\alpha$ at $z=0.4$ and [O{\sc
    iii}]/H$\beta$ at $z=0.83$ for the NB921 emitters.
    
\subsubsection{Spectroscopic redshift analysis} \label{spectro_analysis}

Although an extensive spectroscopic sample is not yet available in the UKIDSS-UDS field (O. Almaini et al. in prep.; H. Pearce et al. in prep.), matching the samples of emitters with all the spectroscopic redshifts published in the literature in this field \citep{Yamada05, Bart_Simpson06, Geach008,van_breu07, Ouchi2008, Smail08, Ono09}\footnote{See UKIDSS UDS website for a redshift compilation by O. Almaini.} allows the
spectroscopic confirmation of 17 H$\alpha$ emitters at $z=1.455-1.48$
(over the full $0.78\deg^2$ area), two [O{\sc iii}]\,5007 emitters at
$z=2.23$, an H$\beta$ emitter at $z=2.33$, and three lower redshift
emitters ([N{\sc i}] at $z=0.545$, [S{\sc viii}]\,9914 at $z=0.625$
and [O{\sc ii}]\,7625 at $z=1.125$), for the NB$_{\rm H}$
data (see Figure~\ref{spec_vs_photo}). Furthermore, follow-up 
spectroscopy with SINFONI on the VLT of 6
H$\alpha$ candidates has resulted in confirming all those sources,
with spectroscopic redshifts $1.45-1.47$ (A. M. Swinbank et al. in prep.), and
observations with FMOS on Subaru have confirmed a further 8 H$\alpha$
emitters (E. Curtis-Lake et al., in prep.), resulting in a total of 31
spectroscopically confirmed $z\sim1.47$ H$\alpha$ emitters.

For NB921 emitters, the limited spectroscopy confirms nine $z=0.4$
H$\alpha$ emitters, 12 $z=0.83$ [O{\sc iii}]\,5007 emitters, a $z=1.1$ H$\gamma$ emitter, ten 4000\AA \ breaks (in which the NB921 filter probed light just to the red of the
break, while the $z'$ band is dominated by emission blueward of the
break, resulting in an excess in NB921), eight [O{\sc ii}] emitters at
$z=1.47$, nine $z=2.31$ MgII emitters (AGN) and a $z=3.87$ CIII] emitter,
  among others.

%
%
\begin{figure}
\centering
\includegraphics[width=8.2cm]{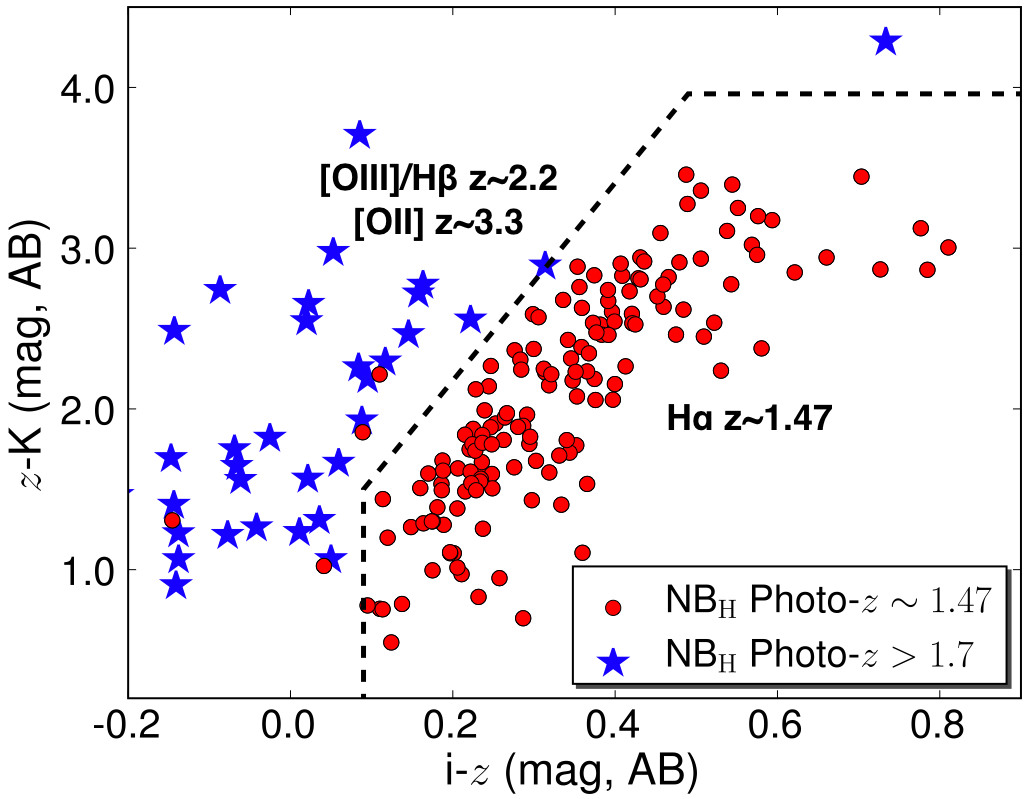}
\caption[Colour-colour separation of higher redshift galaxies]{ The
  $z-$K vs. i$-z$ (AB) colour-colour distribution of NB$_{\rm H}$ emitters,
  demonstrating a method to distinguish between the population of
  $z\sim1.47$ H$\alpha$ emitters, which dominate the sample, and
  higher redshift emitters such as [O{\sc iii}] and H$\beta$ emitters
  at $z\sim2.2$. The dashed line presents the simple separation used
  (in combination with the removal of lower redshift emitters using
  the criteria in Figure~\ref{colour_brik}) for sources with no
  photometric redshifts. The colour separation is similar to the B$z$K
  selection, but by using the $i-z$ colour it better picks up the
  4000\AA \ break for $z=1.47$ galaxies, thus separating them better
  from $z\sim2$ galaxies than using $B-z$. The two pivot points for the $z\sim2.2$ and $z\sim1.47$ separation are [0.09,1.50] and [0.49,3.96]. \label{colour_selection2}}
\end{figure}

\subsubsection{Colour-colour separation of emitters} \label{spectro_analysis}

Colour-colour diagnoses can be valuable tools to explore the extremely
deep broad-band photometry available, particularly for sources with no
photometric redshift information (either because they are faint in the
$K$-band, or because they are found outside the matched Subaru-Spitzer
area). In S09, the $BRiK$ colour-colour diagram ($B-R$ vs. $i-K$) is shown to isolate
$z\sim0.8$ emitters from lower and higher redshift emitters. As
Figure \ref{colour_brik} shows, the same
colour-colour separation is also suited to separate $z\sim1.5$
emitters from lower redshift emitters. Such colour-colour space is
found to be particularly suited to distinguish between the bulk of the
NB921 emitters, as Figure \ref{colour_brik} shows, clearly
isolating H$\alpha$, [O{\sc iii}]/H$\beta$ and [O{\sc ii}] emitters. Based on the
spectroscopic confirmations and the reliable photometric redshifts,
empirical colour-colour selection criteria are defined (see Figure
\ref{colour_brik}) to distinguish between emitters. The same 
separation line produces relatively clean samples of $z>1.3$
emitters for the NB$_{\rm H}$ sample. However, for this filter the
H$\alpha$ emitters at $z\sim1.5$ and the [O{\sc iii}]/H$\beta$ emitters at 
$z\sim2-3$ have similar $BRiK$
colour-colour distributions, and a new set of colours needs to be
explored to separate $z\sim1.5$ and $z>2$ emitters, after using the
$BRiK$ technique to remove low-z emitters. As Figure
\ref{colour_selection2} shows, $i-z$ vs. $z-K$ colours provide a good
separation between $z\sim1.5$ and $z>2$ emitters (the selection is
similar to the widely used $B$$z$$K$ method, but can separate $z\sim1.5$
galaxies from those at $z\sim2$ and higher much more effectively),
defining a colour-colour sub-space where H$\alpha$ emitters
at $z=1.47$ should be found. 

These results demonstrate that a relatively good selection of $z \sim 1.5$
emitters can be obtained using just 5-band ($B$$R$$i$$z'$$K$)
photometry. The $BRiK$ selection can be used to remove low-z emitters,
and a further $i$$z$$K$ analysis is capable of removing higher redshift
emitters (although there's little contamination of the latter in the
NB921 sample).

%
%
\begin{figure}
\centering \includegraphics[width=8.2cm]{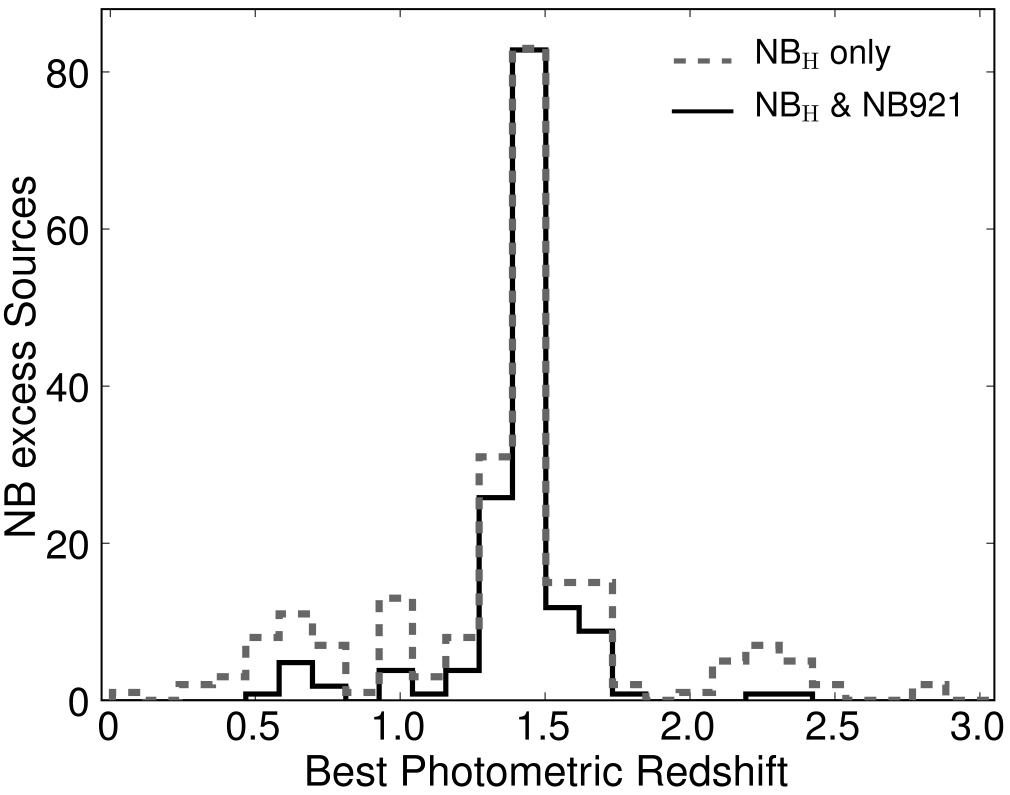}
\caption[Photometric redshift distribution of matched sample of
  emitters]{The photometric redshift distribution of simultaneous
  NB$_{\rm H}$ \& NB921 narrow-band emitters when compared to all
  NB$_{\rm H}$ emitters within the NB921 imaging area. By matching
  both samples, it is possible to define a robust sample of
  H$\alpha$\&[O{\sc ii}] emitters without the need for colour or
  photometric redshift selection. Note that emitters with $K<23$\,AB are not found in the photo-z catalogue, and therefore not shown in the figure. \label{photoz_match}}
\end{figure}

\subsection{The dual narrow-band approach at $z=1.47$} \label{doube_approach}

As Figure \ref{F_profiles} demonstrates, the two narrow-band filter
profiles are extremely well matched in redshift when considering the detection of the H$\alpha$ and [O{\sc ii}] emission lines (although the NB921 filter probes a slightly wider range in redshift). The match can be fully
explored to select robust H$\alpha$ emitters at $z=1.47$, as these should be
detected as [O{\sc ii}] emitters in NB921 (because of the depth of those
data). Within the matched NB$_{\rm H}$-NB921 area (out of a sample of 297 NB$_{\rm H}$ emitters), a sample of 178 dual-emitter sources
is recovered. Figure \ref{photoz_match} shows the photometric redshift
distribution of the double emitters, compared with the distribution of all
NB$_{\rm H}$ emitters with photo-zs (for roughly the same area). The results clearly suggest that the
dual-emitter candidate criteria is able to recover essentially the
entire population of $z=1.47$ H$\alpha$ emitters. Moreover, the dual-emitter
selection also recovers some sources with photometric redshifts which are
significantly higher (3 sources) and lower (14 sources) than $z=1.47$ (accounting for the 1\,$\sigma$ error in the photo-$z$s) and that would have been missed by a simple photo-$z$ selection. There are spectroscopic redshifts available for 7 out of the 14 lower-photo-$z$ sources, and they all confirm those sources to be at $z=1.45-1.48$, indicating that their selection by the dual-emitter technique is very reliable.

The sources for which the photo-$z$ fails have high H$\alpha$ fluxes, and/or are very blue. It is therefore likely that the photo-$z$s fail for these sources because they do not present identifiable breaks (pushing photo-$z$s to a lower redshift solution), and/or because they present strong enough H$\alpha$ lines to contaminate the photometry. 

\subsubsection{Searching for low [O{\sc ii}] EWs H$\alpha$ emitters} \label{double_approach_EW}

Figure \ref{RATIOS} shows the distribution of [O{\sc ii}]/H$\alpha$ line ratios for the emitters
selected in both narrow-band filters. In order to ensure
maximum completeness at the faintest [O{\sc ii}] fluxes, and minimise any
possible biases, a search for extra H$\alpha$ emitters with significant
($\Sigma > 3$) NB921 colour-excess but low [O{\sc ii}] EWs
\ (i.e. waiving the 25\AA \ EW requirement), is conducted. This yields 12
additional sources above the NB921 flux limit. All of these additional sources present
colours and photometric redshifts consistent with being genuine
$z\sim1.47$ sources and probe down to the lowest [O{\sc ii}]/H$\alpha$
line fractions (see Figure \ref{RATIOS}); their colours are consistent
with being sources affected by higher dust extinction. These 12 sources
are added to the robust double-emitter sample.

%
%
\begin{table}
 \centering
  \caption{A summary of the number of sources within the different samples, including detections, candidate emitters, those within the matched area, those for which the full set of colours are available, sources with available photometric redshifts, and those selected as $z=1.47$ emitters.}
  \begin{tabular}{@{}ccc@{}}
  \hline
  Sample  & NB$_{\rm H}$ & NB921  \\
 \hline
   \noalign{\smallskip}
NB Detections ($>3$\,$\sigma$, full area) & 23394 & 347341 \\
Candidate emitters (full area) & 439 & 8865 \\
After visual+star rejection (full area) & 411 & 8747 \\
Matched area NB$_{\rm H}$+NB921 (0.67 sq deg) & 297 & 5505 \\
Selection colours available (0.67 sq deg) & 297 & 2715 \\
Photometric redshifts available & 257 & 1021 \\
Selected $z=1.47$ emitters (0.67 sq deg) & 190 (H$\alpha$) & 1379 ($[\rm OII]$) \\
 \hline
\end{tabular}
\label{numbers}
\end{table}

\subsubsection{The completeness of the H$\alpha$-[O{\sc ii}] double selection} \label{double_approachcomp}

The completeness limit on the [O{\sc ii}]/H$\alpha$ flux ratio (see
Figure~\ref{RATIOS}) indicates that the deep NB921 flux limit guarantees a
very high completeness of the dual-emitter sample, and suggests that the
small number of sources that photo-zs indicate as being at $z\sim1.47$,
but which are not NB921 excess sources, are actually at other redshifts.
To test this further, the photometric redshift and colour-colour
selections are applied in the same matched sky area. These identify 32
H$\alpha$ candidate sources which are not selected by the dual-emitter
approach. Four of these sources have spectroscopic redshifts available,
and all four of these are indeed spectroscopically confirmed to lie at different
redshifts (cf. Figure~\ref{spec_vs_photo}). Of the remaining 28, 18
sources have significantly negative excesses in NB921, and the remaining
10 have $<3\, \Sigma$ excesses, which would correspond to a range of NB921
line fluxes from $\sim0.1-1\times10^{-18}$\,erg\,s$^{-1}$\,cm$^{-2}$.
If real, these line fluxes would imply very low [O{\sc ii}]/H$\alpha$ flux
ratios, indicating that the sources must be highly extinguished (A$_{\rm H\alpha}\sim2-4$\,mag); however,
their UV and optical colours appear inconsistent with such high
extinction (see \S5.5). Thus, these sources are likely to be a
mixture of other line emitters of weaker emission lines close to H$\alpha$
(e.g. [N{\sc ii}], [S{\sc ii}]), which the photometric redshifts are not
sufficiently accurate to distinguish (see Sobral et al. 2009a). 

\subsection{Selecting robust H$\alpha$ and [O{\sc ii}] ${\bf z=1.47}$ emitters} \label{selecting_z147}

\subsubsection{H$\alpha$ emitters at $z=1.47$} \label{selecting_Haz147}

\S~\ref{doube_approach} shows that with the flux limits of this
study, the NB$_{\rm H}$-NB921 is clearly a clean and highly
complete means of selecting $z=1.47$ H$\alpha$ emitters. The sample of robust
H$\alpha$ emitters at $z=1.47$ is therefore derived solely by using the
dual-emitter selection, within the matched area. This results in the
selection of very robust 190 H$\alpha$-[O{\sc ii}] sources at $z=1.47$.

%
%
%
\begin{figure}
\centering
\includegraphics[width=8.2cm]{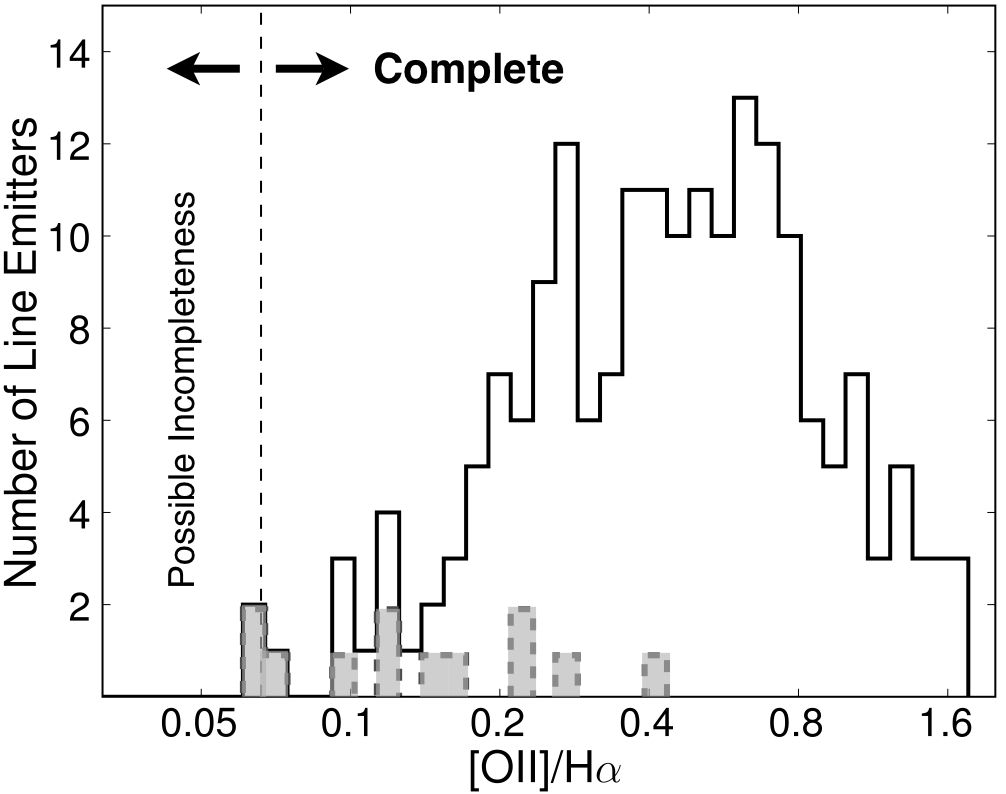}
\caption[Line ratio distributions and survey limits]{The distribution
  of the [O{\sc ii}]/H$\alpha$ line ratios for the robust sample of
  190 simultaneous H$\alpha$-[O{\sc ii}] line emitters at
  $z=1.47$ (the shaded region represents the 12 sources with the low
[O{\sc ii}] EWs). The vertical dashed line represents the lowest line ratio that the
  combined survey is able to probe at the lowest H$\alpha$ fluxes;
  this shows that the [O{\sc ii}] survey is deep enough to recover
  essentially all H$\alpha$ emitters and that the line flux distribution
  is not a result of a bias caused by insufficient sensitivity at the
  lowest line flux ratios.  \label{RATIOS}}
\end{figure}

\subsubsection{[O{\sc ii}] emitters at $z=1.47$} \label{selecting_OIIz147}

In order to select [O{\sc ii}] emitters, in addition to the dual-emitter
selection at bright fluxes, both photo-zs and(or) colours are used. In
particular, the $z=1.47$ [O{\sc ii}] candidates are defined to be narrow-band
excess sources with $z_{min}<1.47<z_{max}$ (where $z_{min}$ and $z_{max}$
are the 3-$\sigma$ redshift limits of the principle peak in the
photometric redshift probability distribution -- photo-zs of emitters have
a typical $\Delta z$ of $\approx0.13$) or those that are found within the
defined $BR-iK$ colour-colour space ($(B-R)<0.26$ or $(i-K)>4.92$ or
$(i-K)>4.121(B-R)+1.349$; see Figure \ref{colour_brik}). Even though the
contamination from higher redshift ($z>1.5$) emitters (such as MgII) is
likely to be low, the colour-colour selection presented in Figure
\ref{colour_selection2} is also applied ($(i-z)<0.24$ or $(z-K)>5.33$ or
$(z-K)>6.15(i-z)+1.394$), to remove potential higher redshift ($z>1.5$)
emitters. This results in a sample of 1379 [O{\sc ii}] selected emitters
down to the flux limit of the survey. However, as noted before, many faint
NB921 emitters are not detected in at least one of the bands necessary for
the colour-colour identification, and therefore are not included directly
in the [O{\sc ii}] sample. \S \ref{OIIcomplet} investigates how this
can bias the determination of the faint-end of the luminosity function and
derives a correction to account for the sources which are missed.

\section{H$\alpha$ and [O{\sc \bf ii}] Luminosity functions} \label{LFs}

\subsection{Contribution from adjacent lines}  \label{cont_adj_lines}

While the NB921 filter can measure the [O{\sc ii}] emission line
(which, in fact, is a doublet) without contamination from any other
nearby emission line, this is not the case for NB$_{\rm H}$ and the H$\alpha$
line. The nearby [N{\sc ii}] emission lines can contribute to the
NB$_{\rm H}$ emission flux, and therefore both EWs and fluxes will be
a sum of [N{\sc ii}] and H$\alpha$. The limited spectroscopic
follow-up in the $H$ band with {\sc sinfoni} (A. M. Swinbank et al. in prep.)
provides good enough S/N and spectral resolution to compute individual
[N{\sc ii}] fractions/corrections. The data show a range of [N{\sc ii}] fractions between
0.1--0.4 and, despite being a limited sample, the results are broadly
consistent with Equation 3 of S09, which estimates the [N{\sc ii}] contamination as a function of total rest-frame EW$_0$([N{\sc ii}]+H$\alpha$), based on a large SDSS sample. For the analysis presented in this paper, the
polynomial approximation used in S09 has been re-computed using a
higher order polynomial which is able to reproduce the full SDSS
relation between the average, $\log$([N{\sc ii}]/H$\alpha)$, $f$, and
$\log$[EW$_0$([N{\sc ii}]+H$\alpha$)], $E$:
$f=-0.924+4.802E-8.892E^2+6.701E^3-2.27E^4+0.279E^5$. This
relation is used to correct all H$\alpha$ fluxes\footnote{The relation
  is used to re-compute H$\alpha$ fluxes in S09 for full consistency;
  the average correction at $z=0.84$ is 0.28 (which compares to 0.25 in S09).} in this paper. The
average correction for the $z=1.47$ sample is 0.22 (average rest-frame
EW$_0$(H$\alpha$+[N{\sc ii}]) of 130\AA).

%
%
\begin{figure}
\centering
\includegraphics[width=8.2cm]{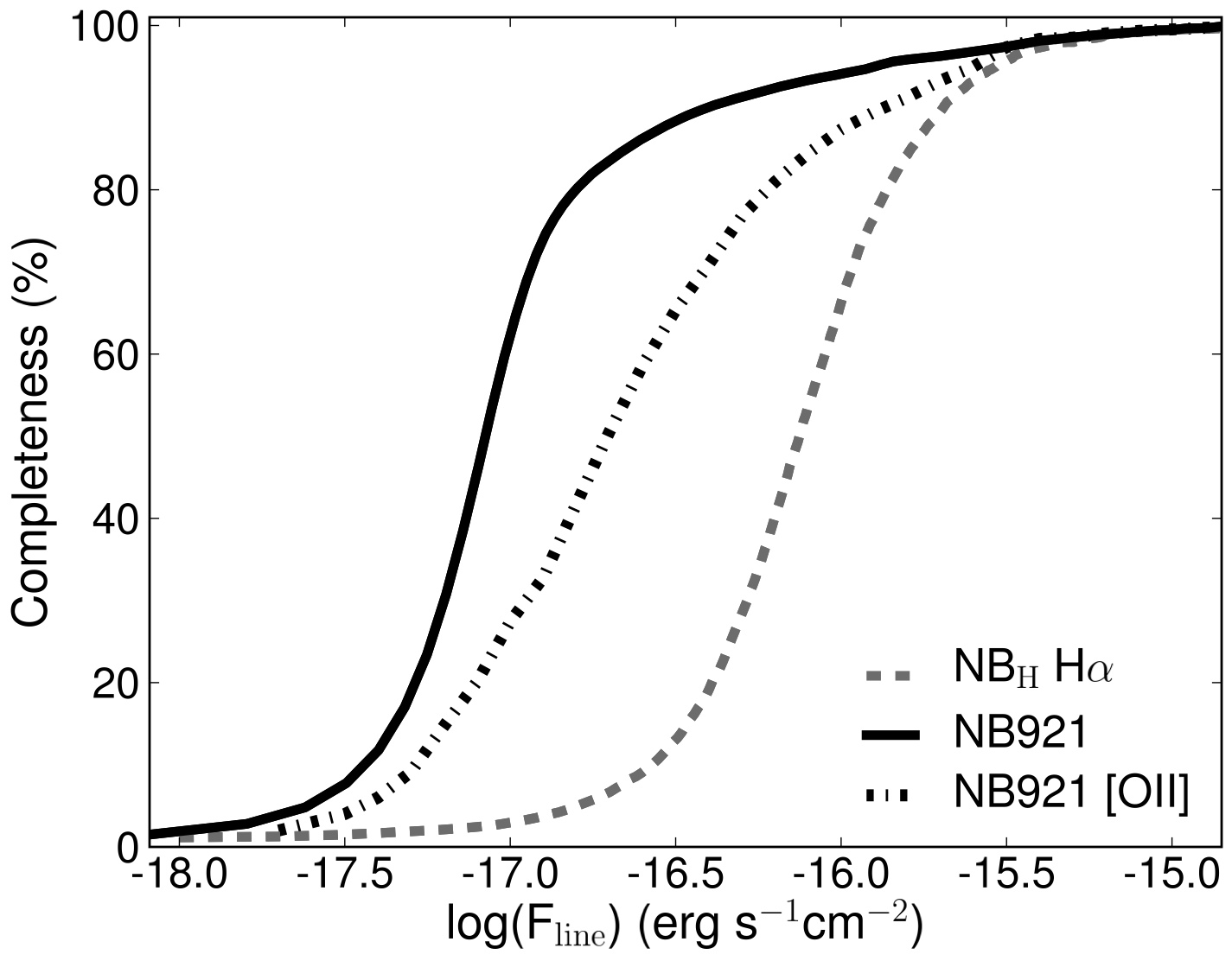}
\caption[Completeness as a function of line flux]{A study of the
  completeness fraction (defined as the fraction of sources with a
  given flux recovered by the selection against the actual number of
  sources with that flux) in order to address incompleteness as a
  function of line flux for both NB$_{\rm H}$ and NB921 samples in the
  SXDF-UKIDSS-UDS field. Note that both image detection and
  colour-magnitude selection are taken into account as these are
  $both$ sources of incompleteness. This confirms the flux limit
  computed for both bands (down to $\sim3\,\sigma$). Note that the NB$_{\rm H}$ curve shown is used to correct the H$\alpha$ luminosity function directly for incompleteness (for each field), while the NB921-[O{\sc ii}] function plotted is the one that is used when correcting the [O{\sc ii}]  luminosity function (obtained by including both the flux incompleteness and the selection incompleteness -- see Figure 11 and Section 4.3). \label{incompleteness}}
\end{figure}

\subsection{Individual line completeness}  \label{complet}

It is fundamental to understand how complete the samples are as
a function of line flux. This is done using simulations, as described
in S09. Briefly, the simulations
consider two major components driving the incompleteness fraction: the
detection completeness (which depends on the actual imaging depth and
the apertures used) and the incompleteness resulting from the
selection (both EW and colour significance). The first component is
studied individually per frame, by adding a set of fake galaxies with
a given input magnitude to each frame and obtaining both the recovery
fraction and the recovered magnitude. The second component is studied
by using sources which have not been selected as emitters, and adding an emission line with a
given flux to all those in order to study the fraction
recovered. In an improvement to S09, sources classed as stars and those occupying the $z<1$
region of the $BRiK$ diagram are not used in this simulation -- this results in a
set of galaxies with an input magnitude distribution which is very
well-matched to the $z=1.47$ population, providing a more realistic
sample to study the line completeness of the survey for $z=1.47$
H$\alpha$ emitters (see Appendix A, which quantifies the difference of
using this method)\footnote{The simulations done in S09 are repeated following this approach, and the results are used to re-compute the H$\alpha$ luminosity function at $z=0.84$ in order to guarantee a completely consistent comparison.}.

For each recovered source, a detection completeness is associated,
based on its new magnitude. The results (survey average) can be found
in Figure \ref{incompleteness}, but it should be noted that because of
the differences in depth, simulations are conducted for each
individual frame, and the appropriate completeness corrections applied
accordingly when computing the luminosity function. 

A similar procedure is followed for the NB921 data (average results
are also shown in Figure \ref{incompleteness}); corrections are also
derived and used individually for each field, although the differences
in depth are not as significant as for the NB$_{\rm H}$
data. Similarly to the NB$_{\rm H}$ analysis, non-emitter sources which are
classed as stars and low redshift ($z<1$) are rejected when studying
how the completeness of the survey varies with line flux in order to
better estimate how complete the survey is specifically for $z=1.47$
line emitters.

For any completeness correction applied, an uncertainty of 20\% of the
size of the correction is added in quadrature to the other
uncertainties to account for possible inaccuracies in the simulations.

%
%
\begin{figure}
\centering
\includegraphics[width=8.2cm]{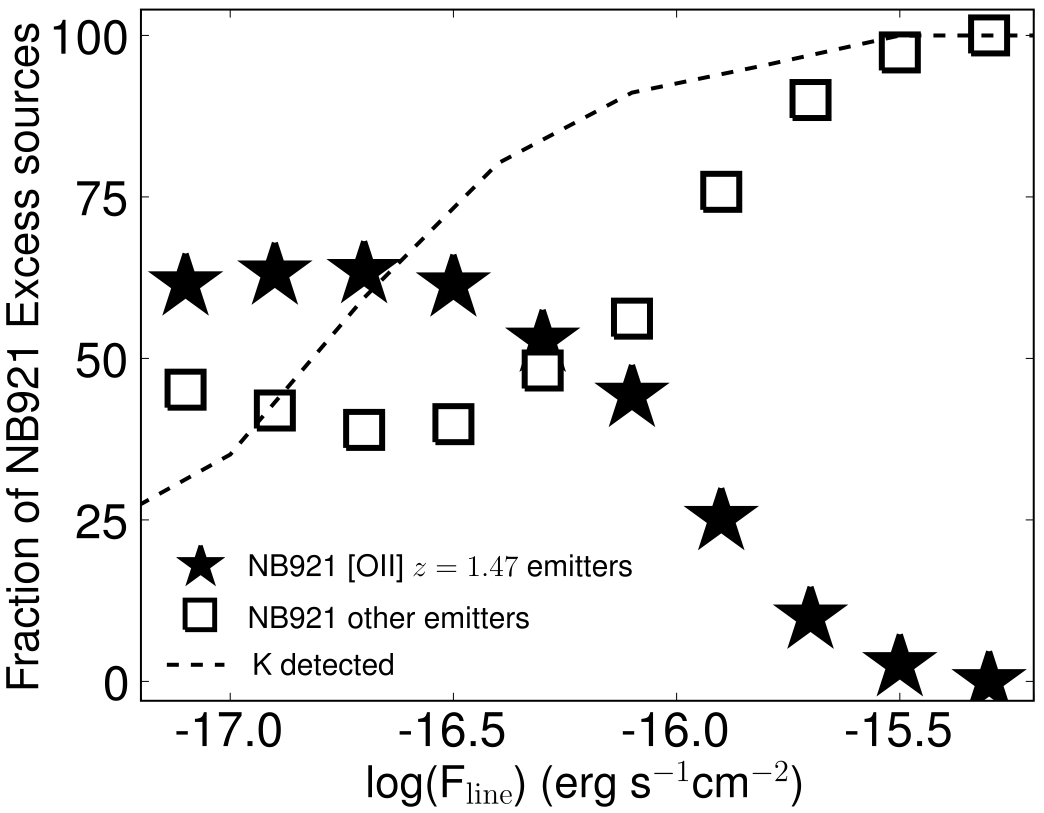}
\caption[Selectability incompleteness]{Fraction of [O{\sc ii}] and
  non-[O{\sc ii}] NB921 emitters (amongst those with $K$ detections, allowing
  for a colour-colour separation) as a function of emission line
  flux. A significant number of potential line emitters are not in the
  final [O{\sc ii}] colour-selected sample as they are not detected in
  the K band imaging; the fraction of all emitters with a $K$
  detection is also shown. \label{line_frac_fluxes}}
\end{figure}

\subsection{[O{\sc ii}] selection completeness}  \label{OIIcomplet}

A significant fraction of NB921-selected emitters do not have
photometric redshifts 
(79\%), and half of the sample is undetected in $K$ above the
$3\sigma$ limit. Therefore, these cannot be robustly classified, and they
are not included in the [O{\sc ii}] sample, even though at least some
of them may well be genuine [O{\sc ii}] emitters at $z=1.47$. This can
cause a potentially significant incompleteness in the [O{\sc ii}]
sample, resulting in an underestimation of the luminosity function. In order
to investigate this source of incompleteness and its effect in the
determination of the luminosity function, the fraction of emitters
with $K$ detections (which generally sets the limit for colour-colour 
selection, as the other bands used are significantly deeper - though 
even in the optical bands typically 33\% of the sources are
undetected) is studied as a function of line flux. As Figure \ref{line_frac_fluxes} shows, the
fraction of NB921 excess sources with $K$ detections falls as a function of decreasing line fluxes,
suggesting that the incompleteness is higher at the lowest
fluxes. This can have implications for determinations of the faint end
slope of the luminosity function.

Figure \ref{line_frac_fluxes} also shows that, for the classifiable sources,  
the fraction of [O{\sc ii}] emitters rises with decreasing flux at the highest 
fluxes, and then seems to remain relatively flat at $\sim60$\% down to 
the lowest fluxes. Assuming that this same distribution is true for
the unclassified sources (which is likely, since the [O{\sc ii}]
fraction also remains roughly constant with $K$--band magnitude for the faintest $K$ magnitudes probed), it is possible to estimate a correction at a given flux that is given by the ratio of the classifiable sources to all sources at that flux. This is equivalent to including unclassified emitters in the luminosity function calculation with a weight that is given by the fraction of [O{\sc ii}] emitters at that flux (see Figure 10). An uncertainty corresponding to 20\% of this correction is
added in quadrature to the other uncertainties.

\subsection{Volume: H$\alpha$ and [O{\sc ii}] surveys}  \label{volume}

Assuming the top-hat (TH) model for the NB$_{\rm H}$ filter (FWHM of
211.1\,\AA, with $\lambda^{TH}_{min}=1.606\,\umu$m and
$\lambda^{TH}_{max}=1.627\,\umu$m), the H$\alpha$ survey probes a
(co-moving) volume of $2.667\times10^5$\,Mpc$^3$. When matched to the
NB921 coverage, this is reduced to $2.2872\times10^5$\,Mpc$^3$ due to the reduction in area -- the matched
FWHM is the same. While this is the total volume probed for the common
flux limit over the entire coverage, each pointing reaches a slightly
different flux limit, and therefore at the faintest fluxes the volume
is smaller (as only one pointing is able to probe those). This is
taken into account when determining the luminosity function: only
areas with a flux limit above the flux (luminosity) bin being
calculated are actually taken into account. In practice, this results
in only using $\sim$ one quarter of the total area for the faintest
bin (with this being derived from the deeper WFCAM pointing).

Assuming a top-hat model for the NB921 filter, (FHWM of 132\,\AA, with
$\lambda^{TH}_{min}=0.9130\,\umu$m and
$\lambda^{TH}_{max}=0.9262\,\umu$m), the [O{\sc ii}] survey probes a
volume (over 0.67\,deg$^2$) of
$2.5102\times10^5$\,Mpc$^3$ when assuming a single line at 3727\AA,
and a volume of $2.6363\times10^5$\,Mpc$^3$ using the fact that the
[O{\sc ii}] line is actually a doublet; 3726.1\AA \ and 3728.8\AA. For
simplicity, because the change in volume is less than 5 per cent (and
therefore much less than the errors), and for consistency with other
authors (allowing a better comparison), the volume used assumes a
single line.

\subsection{Filter Profiles: volume and line ratio biases}  \label{filter_profiles}

Neither NB$_{\rm H}$ or NB921 filter profiles are perfect
top-hats (see Figure \ref{F_profiles}). In order to evaluate the effect of this bias on
estimating the volume (luminous emitters will be detectable over
larger volumes -- although, if seen in the filter wings, they will be
detected as fainter emitters), a series of simulations is
done. Briefly, a top-hat volume selection is used to compute a
first-pass (input) luminosity function and derive the best Schechter
Function fit. The fit is used to
generate a population of simulated H$\alpha$ emitters (assuming they are
distributed uniformly across $1.40<z<1.52$); these are then folded
through the true filter profile, from which a recovered
luminosity function is determined. Studying the difference between the
input and recovered luminosity functions shows that the number of
bright emitters is underestimated, while faint emitters are slightly
overestimated (c.f. S09 for details). This allows correction factors
to be estimated -- these are then used to obtain the corrected
luminosity function.

\begin{figure}
\centering
\includegraphics[width=7.9cm]{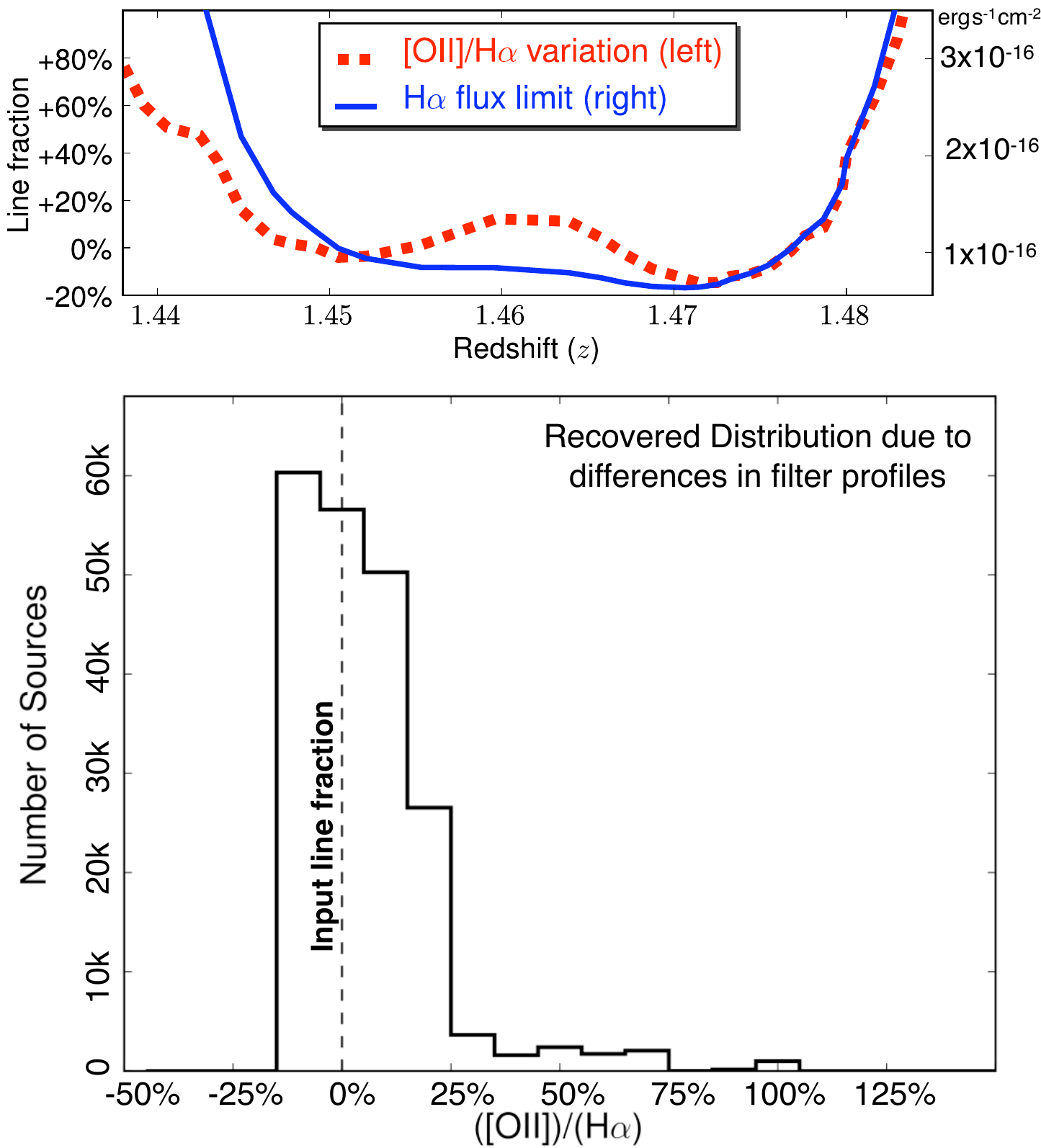}
\caption[A study of the potential profile biases]{$Top$: the variation
  of the measured [O{\sc ii}]/H$\alpha$ line ratio as a function of galaxy
  redshift due to the small differences in the transmission function
  of both filter profiles. Note that the flux limit (in units of
  erg\,s$^{-1}$\,cm$^{-2}$) also varies very significantly as a
  function of redshift, and thus line ratios are likely to be correct to within less than 20\% for the large bulk of the sample. $Bottom$: by
  using the H$\alpha$ luminosity function to produce a population of
  galaxies distributed over a wider redshift range than the
  filter profile, it is possible to use both NB$_{\rm H}$ and NB921 filter
  profiles to study in a simple way the distribution of the ratio
  between the recovered and input line fractions. The results recover the input line fraction with 17\% standard
  deviation and with less than 1\% recovered with a line fraction
  increased by more than +50\%.  \label{biases_filters}}
\end{figure}

Figure \ref{F_profiles} also shows how the NB$_{\rm H}$ and NB921
filters are very well matched, although the [O{\sc ii}] coverage is
slightly wider than the H$\alpha$ coverage. In order to evaluate how this might affect the results on line ratios, 
a series of simulations is done. Simulated [O{\sc ii}]+H$\alpha$ emitters are
distributed uniformly in a volume defined by $1.40<z<1.52$, which
contains the entire transmission regions of both profiles. Emitters
are given a wide range of H$\alpha$ fluxes based on
the observed H$\alpha$ luminosity function, and [O{\sc ii}] fluxes 
corresponding to [O{\sc
    ii}]/H$\alpha$ line ratios between 0 and 2.0. The real filter
profiles are then used to
recover, for each emitter, both the [O{\sc ii}] and H$\alpha$ fluxes,
and therefore allow the study of both the recovered line fluxes and
the recovered line ratios. Based on these results -- presented in
Figure \ref{biases_filters} -- the line ratios should be accurate
within $\sim20$\%.

%
%

\subsection{Extinction Correction}  \label{ext_corr}

The H$\alpha$ emission line is not immune to dust extinction, although
it is considerably less affected than the [O{\sc ii}] emission
line. Measuring the extinction for each source can in principle be
done by several methods, ranging from spectroscopic analysis of Balmer
decrements to a comparison between H$\alpha$ and far-infrared
determined SFRs, but such data are currently not available. 

In this Section, the analysis of the H$\alpha$ luminosity function is
done assuming A$_{\rm H\alpha}=1$\,mag of extinction at
H$\alpha$\footnote{Corresponding to $\sim1.76$\,mag of extinction, or a factor of 5$\times$ at
  [O{\sc ii}] for a Calzetti et al. (2000) extintion law.}, as this
allows an easy comparison with the bulk of other studies which have
used the same approach. A few studies have used a H$\alpha$ dependent
extinction correction -- either derived from \cite{Garn2010a}, or from \cite{Hopkins}. However, \S \ref{OII_Ha_view}
suggests that at least the overall normalization of such relation at
$z\sim1.5$ is significantly lower (dust extinction is not as high as
predicted for the very high luminosities probed) and therefore
over-predicts dust-extinction corrections by
$\sim0.5-0.7$\,mag (overcorrecting luminosities by $0.2-0.3$ in dex). Finally, the [O{\sc ii}] luminosity function is
presented without any correction for extinction, in order to directly
compare it with the bulk of other studies. Detailed extinction
corrections and a discussion regarding those are presented in Section
\ref{OII_Ha_view}.


\subsection{H$\alpha$ Luminosity Function at $\bf z=1.47$ and Evolution}  \label{LF_Ha}

By taking all H$\alpha$ selected emitters, the luminosity function is
computed. As previously described, it is firstly assumed that the
NB$_{\rm H}$ filter is a perfect top-hat, but the method of S09 is
applied to correct for the real profile (see Section
\ref{filter_profiles}). Candidate H$\alpha$ emitters are assumed to be
all at $z=1.47$ (as far as luminosity distance is concerned). Results
can be found in Figure \ref{LF_Halpha_HALPHA}. Errors are Poissonian, with a further 20\% of completeness corrections added in quadrature. The luminosity function
is fitted with a Schechter function defined by three parameters
$\alpha$, $\phi ^*$ and $L^*$:

%
%
\begin{figure}
\centering
\includegraphics[width=8.2cm]{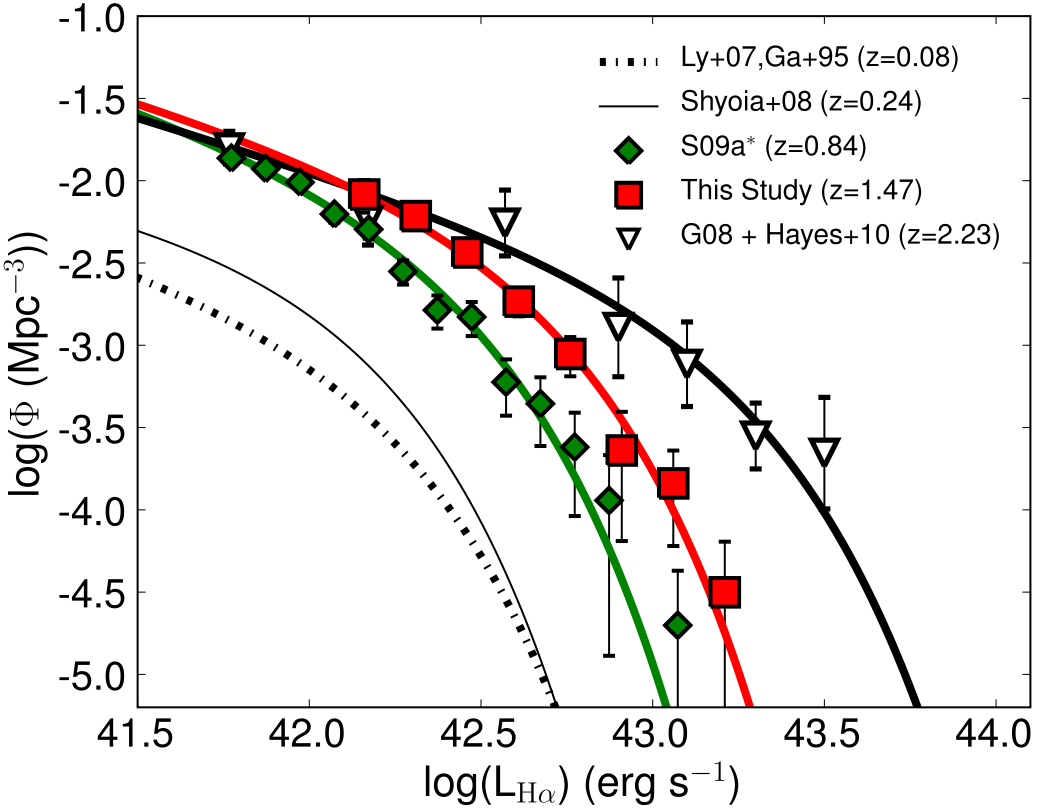}
\caption[The H$\alpha$ luminosity function evolution]{The H$\alpha$
  luminosity function evolution up to $z\sim2.3$. The new measurement at $z=1.47$ is combined with
  HiZELS measurements (S09 and G08) and other narrow-band
  studies \citep{Gallego95,Ly2007,Shioya,G08,Hayes,Tadaki}. This
  confirms the strong L$_{\rm H\alpha}^*$ evolution from $z=0$ to
  $z=2.23$. \label{LF_Halpha_HALPHA}}
\end{figure}

\begin{equation}
  \phi(L) \rm dL = \it \phi^* \left(\frac{L}{L^*}\right)^{\alpha} e^{-(L/L^*)} \rm d\it\left(\frac{L}{L^*}\right).
\end{equation}
In the $\log$ form\footnote{$\log_{10}$; note that the definition of
  $\phi^*$ used in S09 and Geach et al. (2008), $\phi^*_S$, is given
  by $\phi^*_S=\phi^*\times\ln10$.} the Schechter function is
given by:

\begin{equation}
  \phi(L) \rm dL = \ln10 \, \it \phi^* \left(\frac{L}{L^*}\right)^{\alpha} e^{-(L/L^*)} \left(\frac{L}{L^*}\right)\rm d\log L.
\end{equation}
A Schechter function is fitted to the H$\alpha$ luminosity function,
with the best fit resulting in:

\medskip
\noindent $\log L^*_{\rm H\alpha}=42.50\pm0.23 $\,erg\,s$^{-1}$

\smallskip
\noindent $\log\phi^*_{\rm H\alpha}=-2.44\pm0.33 $\,Mpc$^{-3}$

\smallskip
\noindent $\alpha_{\rm H\alpha}=-1.6\pm0.4$
\smallskip

\noindent The best-fit function is also shown in Figure
\ref{LF_Halpha_HALPHA}, together with the luminosity functions
determined by \cite{Ly2007} -- extending the work by \cite{Gallego95}
-- at $z\approx0$ and that from \cite{Shioya} at $z=0.24$. Also presented on Figure \ref{LF_Halpha_HALPHA} are luminosity functions from the other HiZELS redshifts. The new luminosity function uses the revised
catalogues presented in Sobral et al. (2010) and Sobral et al. (2011), and uses new completeness
corrections, re-computed fluxes and the rest-frame EW dependent [N{\sc
    II}] correction. The changes to the luminosity function are only minor. At $z=2.23$, the luminosity function has also been re-calculated by combining the results from \cite{G08}, with new ultra-deep measurements of the faint end by \cite{Hayes} -- using HAWK-I on VLT -- and \cite{Tadaki} -- using MOIRCS on Subaru. Best fits for the HiZELS luminosity
functions at $z=0.84$, $z=1.47$ and $z=2.23$ are presented in Table \ref{lfs__}.

The results confirm the strong evolution from the local Universe to
$z=2.23$ and provide further insight. In particular, while there is
significant evolution up to $z\sim0.8$, the H$\alpha$ luminosity
functions at $z\sim0.8$, $z\sim1.47$ and $z\sim2.23$ \citep[including
  the ultra-deep measurement by][]{Hayes}
seem to agree well at the lowest luminosities probed (below
$\log L_{\rm H\alpha} \sim42.0$) in normalisation and slope -- all consistent with a
relatively steep value of $\alpha\sim-1.6$) -- a similar
$\alpha$ is found by \cite{CHU11}. In contrast, at the bright end,
L$_{\rm H\alpha}^*$ is clearly seen to continue to increase from
$z=0.84$ to $z=1.47$ and to $z=2.23$, with $\log\,L^*\propto0.6z$. This implies that the bulk of
the evolution from $z=2.23$ to $z\sim1$ is happening for the most
luminous H$\alpha$ emitters (L$_{\rm H\alpha}>10^{42}$\,erg\,s$^{-1}$), which
greatly decrease their number density as the Universe ages: 
the faint-end number densities seem to remain relatively unchanged.

\medskip

%
%
\begin{table*}
 \centering
  \caption{The luminosity function and star-formation rate density
    evolution at the peak of the star formation history as seen by
    HiZELS at $z=0.84$, $z=1.47$ and $z=2.23$; assuming 1 mag
    extinction at H$\alpha$. Columns present the redshift, break of
    the luminosity function, $L^*_{H\alpha}$, normalisation,
    $\phi^*_{H\alpha}$ and faint-end slope of the luminosity function,
    $\alpha$. The three right columns present the star formation rate
    density at each redshift based on integrating the luminosity
    function down to the given luminosity limit (in $\log$
    erg\,s$^{-1}$). Star formation rate densities include a correction for AGN contamination of 10\% at $z=0.84$ (c.f. Garn et al. 2010) and 15\% at both $z=1.47$ and $z=2.23$.}
  \begin{tabular}{@{}cccccccccc@{}}
  \hline
   Epoch & $L^*_{H\alpha}$ & $\phi^*_{H\alpha}$ & $\alpha_{H\alpha}$ & $\rho_{SFR H\alpha}$ 42  & $\rho_{SFR H\alpha}$ 40 & $\rho_{SFR H\alpha}$ All  \\
     (z)    & erg\,s$^{-1}$ & Mpc$^{-3}$ &  & M$_{\odot}$\,yr$^{-1}$ Mpc$^{-3}$  & M$_{\odot}$\,yr$^{-1}$ Mpc$^{-3}$  & M$_{\odot}$\,yr$^{-1}$ Mpc$^{-3}$  \\
 \hline
   \noalign{\smallskip}
$z=0.84\pm0.02$ & $42.25\pm0.16$ & $-2.36\pm0.29$ & $-1.66\pm0.34$ & $0.03\pm0.01$ & $0.12\pm0.04$  & $0.13\pm0.06$ \\
\bf $z=1.47\pm0.02$ & $42.50\pm0.23$ & $-2.44\pm0.33$ & $-1.6\pm0.4$ & $0.06\pm0.02$ & $0.15\pm0.06$  & $0.16\pm0.05$ \\
 $z=2.23\pm0.02$ & $43.07\pm0.24$ & $-2.93\pm0.41$ & $-1.60\pm0.21$ & $0.12\pm0.06$ & $0.20\pm0.09$  & $0.22\pm0.10$ \\
 \hline
\end{tabular}
\label{lfs__}
\end{table*}

\subsection{The [O{\sc ii}] Luminosity function at $\bf z=1.47$ and Evolution}  \label{OII_LF}

\begin{figure}
\centering
\includegraphics[width=8.2cm]{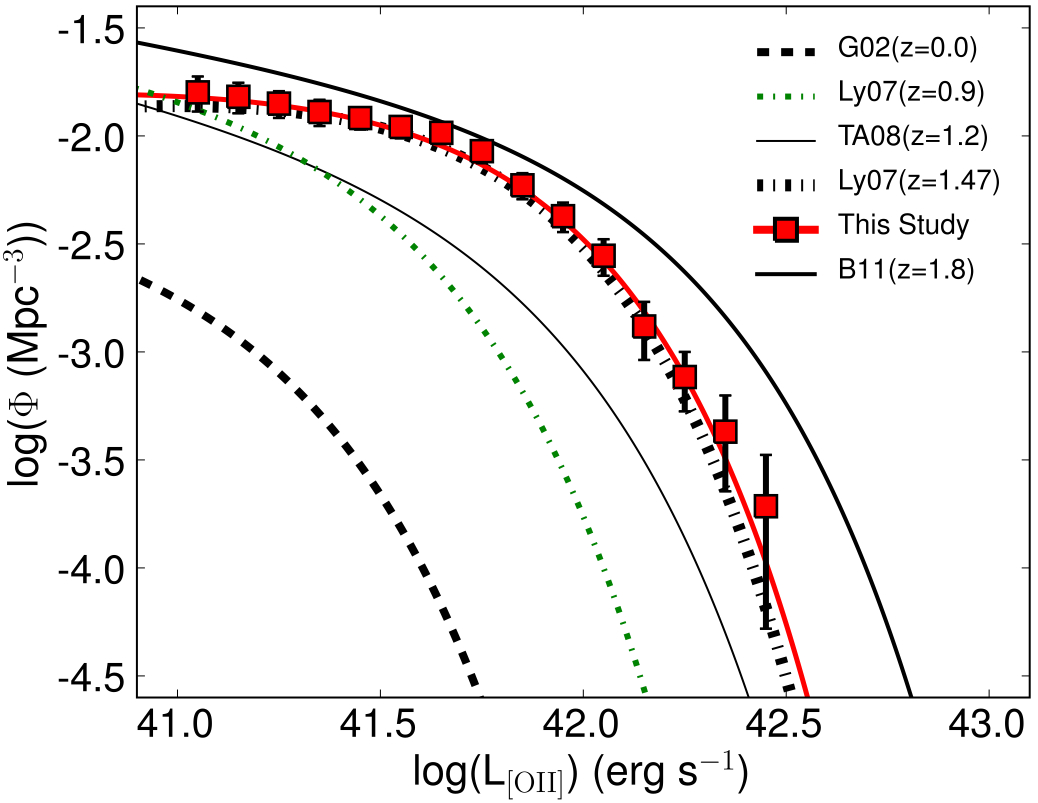}
\caption[The evolution of the OII luminosity function]{The
  [O{\sc ii}] luminosity function derived from this study, at $z=1.47$
  and a comparison with measurements at different redshifts from
  \cite{GalegoOII,Takahashi,Ly2007,Bayliss} and with a measurement
  at the same redshift (using the same narrow-band filter, but on a
  different field) by Ly et al. (2007). There is a clear evolution in
 the [O{\sc ii}] luminosity function, at least up to $z\sim2$. \label{LF_OII}}
\end{figure}

The [O{\sc ii}] luminosity function at $z=1.47$ (not corrected for
dust extinction) is shown in Figure \ref{LF_OII}. Note that this
includes corrections for i) incompleteness in both detection and flux
selection, ii) incompleteness in the redshift selection of [O{\sc ii}] emitters (see Figure 10 for the combined correction) and iii) correction for the filter profile not being a perfect tophat. Errors are Poissonian and include (in quadrature) 20\% of the completeness correction applied).
The fully-corrected [O{\sc ii}] luminosity
function is found to be well-fitted by a Schechter
function, with the best fit resulting in:

\medskip
\noindent $\log L^*_{\rm [O{\sc II}]}=41.71\pm0.09$\,erg\,s$^{-1}$

\smallskip
\noindent $\log\phi^*_{\rm [O{\sc II}]}=-2.01\pm0.10$\,Mpc$^{-3}$

\smallskip
\noindent $\alpha_{\rm [O{\sc II}]}=-0.9\pm0.2$

\smallskip

%
%

The results are compared with other studies at different redshifts
\citep{Hogg,GalegoOII,Takahashi,Ly2007,Bayliss} -- all uncorrected for
dust extinction -- and shown in Figure \ref{LF_OII}. The comparison
reveals a significant evolution in the [O{\sc ii}] luminosity function
from $z\approx0$ to $z=1.47$. Such evolution seems to be simply
described by a $\phi^*$ and L$^*$ evolution up to $z\sim0.9$ and a
continuing L$^*$ evolution from $z\sim0.9$ up to $z\sim2$, in line
with the results for the evolution of the H$\alpha$ luminosity function. It should
also be noted that there is good agreement with the \cite{Ly2007}
luminosity function at the same redshift derived from a smaller area,
but at a similar depth.

The correction for redshift/emitter selection is found to be
particularly important at faint fluxes, setting the slope of the
faint-end of the luminosity function (although it has little effect on
the values of $L^*$ and $\phi^*$). If no correction for unclassified
sources is applied, the best-fit Schechter function yields
$\alpha\approx-0.2$ (compared to the corrected best-fit of
$\alpha\approx-0.9$). In the extreme case where all the
non-selectable emitters are assumed to be [O{\sc ii}], it would yield
$\alpha\approx-1.8$. This makes it clear that the crucial data that
one needs in order to improve the determination of the faint-end slope
is not new, deeper NB921 data, but rather significantly deeper $B$,
$i$, $z$ and $K$ data, which will enable to completely distinguish between different line emitters at faint fluxes (or at least provide a more robust correction by
classifying a much higher fraction of emitters).

%
%
\begin{figure*}
\centering
\includegraphics[width=13.1cm]{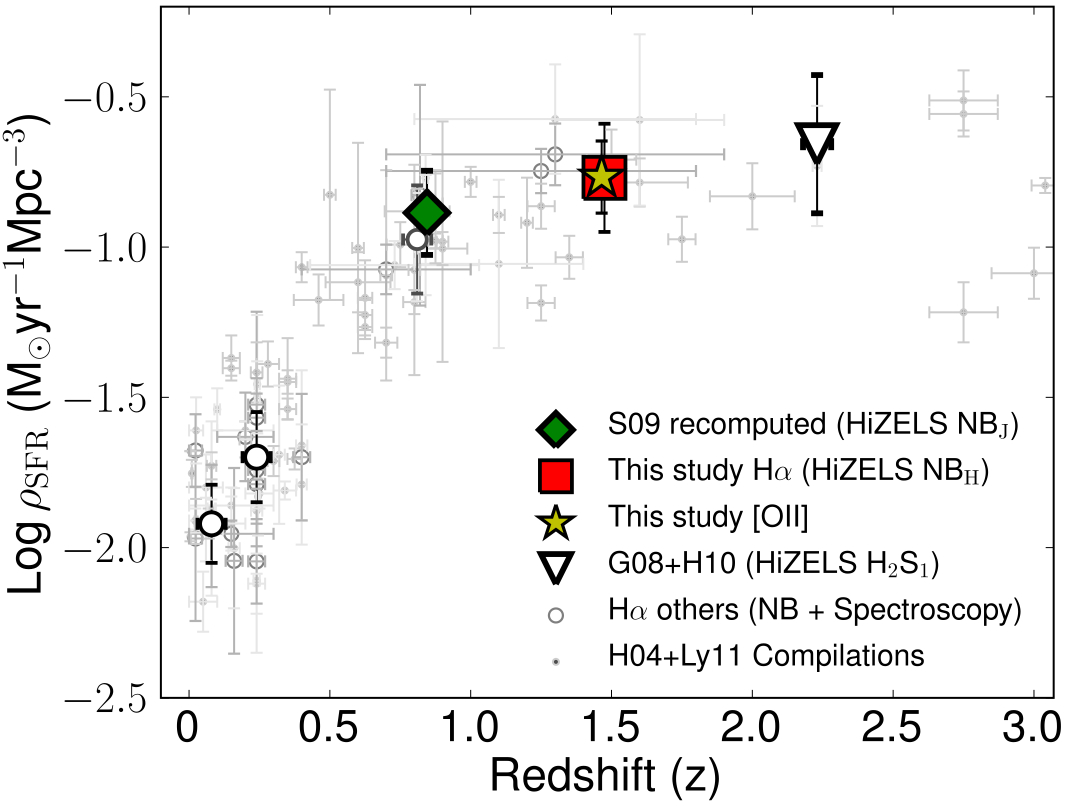}
\caption[SFRD history]{The star formation rate density and its
  evolution with redshift up to $z\sim2.3$ using H$\alpha$ only, but also including the [O{\sc ii}] measurement presented in this paper (assuming 1 magnitude of extinction at H$\alpha$) and compared to estimates at different redshifts from the literature \citep[e.g.][and references herein]{Hopkins2004,CHU11}. Darker circles represent the results from H$\alpha$ studies which are also shown in Figure 13. This confirms a strong evolution in the star formation rate density over the last $\sim10$\,Gyrs, with a flattening or slow decline over $1<z<2$ and a sharp decrease from $z\sim1$ to $z\sim0$. \label{SFRD_lum_lim_Ha}}
\end{figure*}

\subsection{Evolution of the H$\alpha$ and [O{\sc ii}] Luminosity Functions}  \label{view_peak_SFH}

Robust measurements of the evolution of both the H$\alpha$ and [O{\sc
    ii}] luminosity functions up to $z\sim2$ have been presented. The
results reveal that there is a strong, and consistent, evolution of
both luminosity functions. Moreover, while the evolution up to
$z\sim1$ can be described as both an increase of the typical
luminosity (L$^*$) and an increase of the overall normalization of the
luminosity functions ($\phi^*$), at $z>1$ the bulk of the evolution
can be simply described as a continuous increase in L$^*$ given by $\log\,L^*\propto0.6z$. The current
results also show that the faint-end slope (and the number density of
the faintest star-forming galaxies probed) seems to remain relatively
unchanged during the peak of the star formation history
($z\sim1-2$). Of course, the latter does not imply, at all, that the
faint population is not evolving. Furthermore, particularly beyond
$z\sim1$, AGN contamination at the highest luminosities could still be
polluting the view of the evolution of the star-forming population.

\subsection{The star formation rate density at $\bf z=1.47$}  \label{SFRD_z147}

The best-fit Schechter function fit to the H$\alpha$ luminosity
function can be used to estimate the star formation rate density at
$z=1.47$. The standard calibration of Kennicutt (1998) is used to
convert the extinction-corrected H$\alpha$ luminosity to a star
formation rate:
\begin{equation}
{\rm SFR}({\rm M}_{\odot} {\rm yr^{-1}})= 7.9\times 10^{-42} ~{\rm L}_{\rm H\alpha} ~ ({\rm erg\,s}^{-1}),
\end{equation}
which assumes continuous star formation, Case B recombination at $T_e
= 10^4$\,K and a Salpeter initial mass function ranging from
0.1--100\,M$_{\odot}$. A constant 1 magnitude of extinction at H$\alpha$ is assumed
for the analysis, as this is a commonly adopted approach which is a 
good approximation at least for the local Universe. \S
\ref{OII_Ha_view} will investigate this further based on the analysis 
of emission-line ratios, revealing that A$_{\rm H\alpha}\sim1$\,mag for the sample presented in this paper. A 15\% AGN contamination is also assumed (Garn et al. 2010
found AGN $\sim10$\% contamination at $z=0.84$, but the contamination
is likely to be higher at higher redshift and the slightly higher flux
limit of this sample). Down to the survey limit (after extinction correction,
L$_{\rm H\alpha}=10^{42}$erg\,s$^{-1}$), one finds $\rho_{\rm
  SFR}=0.07\pm0.01$\,M$_{\odot}$\,yr$^{-1}$ Mpc$^{-3}$;
an integration extrapolating to zero luminosity yields $\rho_{\rm SFR}=0.16\pm0.05$\,M$_{\odot}$\,yr$^{-1}$ Mpc$^{-3}$. Furthermore, by using the standard Kennicutt (1998) calibration of [O{\sc ii}] as a star formation tracer (calibrated using H$\alpha$):

\begin{equation}
{\rm SFR}({\rm M}_{\odot} {\rm year^{-1}})= 1.4\times 10^{-41} ~{\rm L}_{\rm [OII]}~ ({\rm ergs\,s}^{-1}),
\end{equation}
it is also possible to derive an [O{\sc ii}] estimate of the star formation rate density at the same redshift, by using the complete integral of the [O{\sc ii}] luminosity function determined at $z=1.47$, and assuming the same 1 magnitude of extinction at H$\alpha$. This yields $\rho_{\rm SFR}=0.17\pm0.04$\,M$_{\odot}$\,yr$^{-1}$ Mpc$^{-3}$, in very good agreement with the measurement obtained from H$\alpha$.

By taking advantage of the other HiZELS measurements, and other
H$\alpha$ based measurements, it is possible to construct a full and
consistent view of the H$\alpha$-based star formation history of the
Universe. This is done by integrating all derived luminosity functions
down to a common luminosity limit of L$_{\rm
  H\alpha}$=10$^{41.5}$erg\,s$^{-1}$. The results are presented in Figure
\ref{SFRD_lum_lim_Ha}, revealing a strong rise in the star formation
activity of the Universe up to $z\sim1$ and a flattening or a small
continuous increase beyond that out to $z>2$.

\section{The matched H$\alpha$-[O{\sc ii}] view}  \label{OII_Ha_view}

This section presents a detailed comparison of [O{\sc ii}] and H$\alpha$ luminosities and line flux ratios (observed and corrected for
dust-extinction) as a function of galaxy colour, mass and luminosity for the sample of matched H$\alpha$ and [O{\sc ii}] emitters at $z=1.47$. These are compared with equivalent results for a similarly-selected sample at $z\sim0.1$ drawn from the SDSS.

\subsection{A SDSS sample at $\bf z\sim0.1$}  \label{calib_fractions}

In order to compare the $z=1.47$ H$\alpha$+[O{\sc ii}] sample with a large local sample and provide a further insight into any important correlations between line fractions and dust-extinction, mass and colour, an SDSS sample was used. Data were extracted from the MPA SDSS derived data products catalogues\footnote{See http://www.mpa-garching.mpg.ed/SDSS/}. The sample was defined to emulate a narrow-band slice at $0.07 < z < 0.1$ (chosen to be distant enough so that the fibers capture the majority of the light -- typically $\sim3-4$\,kpc diameter, but not too far out in redshift to guarantee that the sensitivity is still very high and the measurements are accurate). The sample was selected by further imposing a requirement that L$_{H\alpha}>10^{40.6}$\,erg\,s$^{-1}$ (observed, not aperture or dust extinction corrected); this guarantees high S/N line ratios and that the vast majority of sources
are detected in [O{\sc ii}] as well, allowing unbiased estimates of the line ratio distribution, as well as the detection of other
emission lines (O{\sc iii}], H$\beta$ and [N{\sc ii}]) that can be used to distinguish
between AGN and star-forming galaxies \citep[c.f. ][]{Rola,Brinchmann}. Among 17354 SDSS $z\sim0.1$ H$\alpha$
emitters, 498 were classified as AGN, implying a $\sim3$ per cent contamination. Potential AGN were removed from the sample. Emission line fluxes are aperture corrected following a similar procedure as in Garn \& Best
(2010) -- i.e., by using the ratio between the fiber estimated mass
and the total mass of each galaxy. For 32 galaxies where the catalogued fiber
mass is higher than the catalogued total mass, an aperture correction
factor of 1.0 was assigned. For those in which the fiber mass had not been determined,
the average correction for the total mass of that galaxy was
assigned. The median fraction of flux within the SDSS apertures is
32\%. Note that the even though these aperture corrections change the total
fluxes, line ratios remain unchanged. Finally, in order to provide a more direct comparison with the sample at $z=1.47$, an EW cut of 20\,\AA \ in H$\alpha$ was applied (to mimic the selection done at $z=1.47$), and galaxies with lower EWs were excluded (1701 galaxies). The final SDSS sample contains 14451 star-forming H$\alpha$ selected galaxies at $0.07 < z < 0.1$.

\subsection{Calibrating [O{\sc ii}]/H$\alpha$ line ratio as a dust extinction probe}  \label{calib_fractions}

Due to the difference in rest-frame wavelength of the [O{\sc ii}] and H$\alpha$ emission lines, and both emission-lines being tracers of recent star-formation, the [O{\sc ii}]/H$\alpha$ line ratio is sensitive to dust-extinction, even though metallicity can affect the
line ratio as well. After correcting for dust extinction, several studies find a [O{\sc ii}]/H$\alpha$ average of $\approx1.0-1.4$ in the local Universe \citep[c.f.][]{Kewley04}.

The H$\alpha$/H$\beta$ line ratio (Balmer decrement) is widely used as an extinction estimator, particularly up to $z\sim0.4$, as it is relatively easy to obtain both emission lines. As the
SDSS-derived sample is able to obtain reliable fluxes for the [O{\sc ii}], H$\beta$ and H$\alpha$ emission lines, it is possible to
investigate (and potentially calibrate) [O{\sc ii}]/H$\alpha$ as a
dust-extinction indicator, using the Balmer decrement directly. As 
Figure \ref{LHa_LOII_balmer} shows, [O{\sc ii}]/H$\alpha$ is
relatively well correlated with H$\alpha$/H$\beta$, indicating that it
is possible to use [O{\sc ii}]/H$\alpha$ to probe dust extinction
within the observed scatter. For each
galaxy in the SDSS-derived sample at $z\sim0.1$, H$\alpha$/H$\beta$
line fluxes are measured and used to estimate the extinction at
H$\alpha$, A$_{\rm H\alpha}$, by using:

%
%
\begin{figure}
\centering
\includegraphics[width=8.2cm]{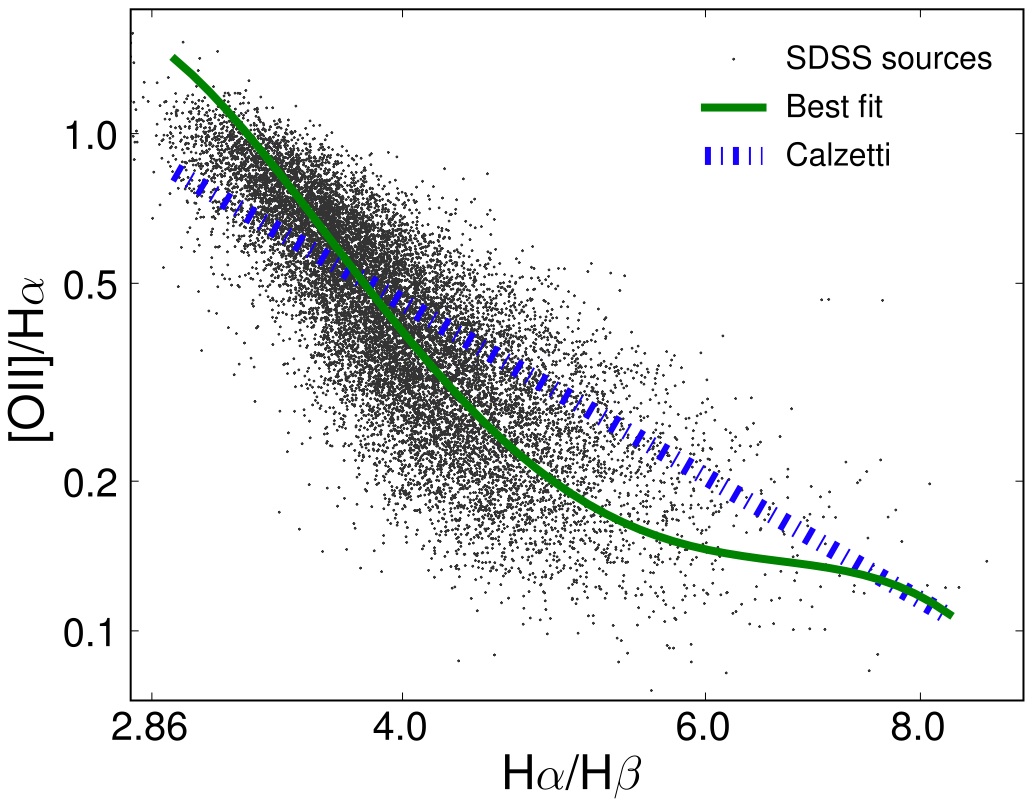}
\caption[Calibrating OII/H$\alpha$ line ratios as dust extinction probes using the Balmer decrement]{The variation of
  [O{\sc ii}]/H$\alpha$ line ratios as a function of H$\alpha$/H$\beta$ ratios for SDSS (balmer decrement), showing that they correlate well, and therefore it is possible to calibrate the observed [O{\sc ii}]/H$\alpha$ line ratio as a dust extinction indicator. The best polynomial fit to the data is also shown, together with the prediction from the Calzetti et al. extinction law. \label{LHa_LOII_balmer}}
\end{figure}

\begin{equation}
A_{\rm H\alpha} = \frac{-2.5 k_{\rm H\alpha}}{k_{\rm H\beta}-k_{\rm
    H\alpha}}~{\rm log}_{10}\left(\frac{2.86}{{\rm H\alpha}/{\rm
    H\beta}}\right),
\end{equation}
where 2.86 is the assumed intrinsic H$\alpha$/H$\beta$ line flux
ratio, appropriate for Case~B recombination, temperature of $T =
10^{4}$~K and an electron density of $n_{\rm e} = 10^{2}$~cm$^{-3}$
\citep{Brocklehurst71}. The \citet{Calzetti00} dust attenuation law is
used to calculate the values of $k_{\lambda} \equiv A_{\lambda} /
E(B-V)$ at the wavelengths of the H$\alpha$ and H$\beta$ emission
lines, resulting in:
\begin{equation}
A_{\rm H\alpha} = 6.531\log_{10} \rm H\alpha/H\beta-2.981.
\end{equation}
By using the Calzetti law, it is also possible to write a similar relation by using [O{\sc ii}]/H$\alpha$:
\begin{equation}
A_{\rm H\alpha} = -2.296\log_{10} \rm [OII]/H\alpha+2.29\log_{10}(R_0), 
\end{equation}
where $R_0$ is the unknown intrinsic [O{\sc ii}]/H$\alpha$, but the slope of the relation is fully determined. Figure \ref{LHa_LOII_balmer} shows that the Calzetti law matches the global trend well, but fails to predict the fine details. The failure to match the data perfectly results from a combination of different factors. The Calzetti et al. law does not include important variations of the intrinsic [O{\sc ii}]/H$\alpha$ line fraction with e.g. metallicity. The Calzetti et al. extinction law is also based on a relatively small number of local galaxies, and, most of all, it is based on continuum light, and not on emission-lines. However, it is possible to obtain a much better fit to the relation between [O{\sc ii}]/H$\alpha$ ($\Psi$) and Balmer decrement with a 4th order polynomial (shown in Figure 16), and to derive an empirical relation between A$_{\rm H\alpha}$ directly from H$\alpha$/H$\beta$:
\begin{equation}
\rm A_{\rm H\alpha}=-4.30\Psi^4-11.30\Psi^3-7.39\Psi^2-2.94\Psi+0.31
\end{equation}
with a scatter of $0.28$ mag (in A$_{\rm H\alpha}$). The scatter may be driven by the range of metallicities within the sample, as galaxies will have slightly different intrinsic [O{\sc ii}]/H$\alpha$ line ratios.

%
%
\begin{figure}
\centering
\includegraphics[width=8.2cm]{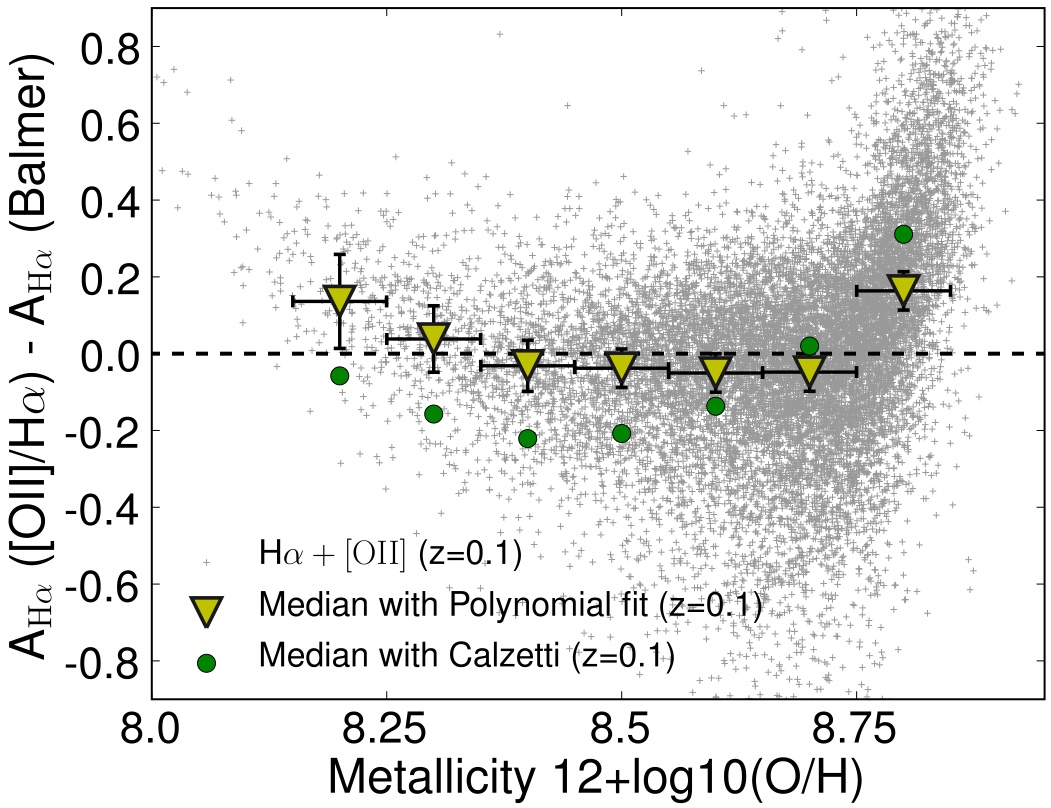}
\caption[Evaluating potential metallicity trends and biases in the calibration]{The difference between A$_{\rm H\alpha}$ estimated using equation 12 and directly using the Balmer decrement, as a function of a metallicity (from Equation 13). The results show that the empirical calibration derived is able to use [O{\sc ii}]/H$\alpha$ line ratios as dust extinction probe without any significant bias for a relatively wide range of metallicities/abundances. A comparison with what would be obtained using the Calzetti law directly is also shown, highlighting that the polynomial fit is able to provide a calibration of [O{\sc ii}]/H$\alpha$ which is less affected by metallicities. Solar metallicity corresponds to 12+$\log10$ of 8.66; the figure presents metallicities varying from $\approx0.2$ to $\approx2.2$ solar metallicities. Note that while Equation 12 provides reliable results (without any significant systematics/biases) for galaxies with abundances (12+$\log10$) from 8.2 to 8.75, it will overestimate the median A$_{\rm H\alpha}$ by about +0.2 mag for very sub-solar and very super-solar metallicity galaxies. \label{MetalDEP}}
\end{figure} The SDSS data can be used to investigate how the offset of a galaxy from the  [O{\sc ii}]/H$\alpha$ vs H$\alpha$/H$\beta$ relation depends upon metallicity, and thus it is possible to measure the extent to which the [O{\sc ii}]/H$\alpha$ line ratio can be used to probe extinction, without metallicity biases. Here, the O3N2 indicator \citep{PettiniPagel04} is used as a tracer of metallicity (the gas-phase abundance of oxygen relative to hydrogen), computed by using:
\begin{equation}
\rm 12+\log10(O/H)=8.73-0.32\times\log_{10}(O3H\beta/N2H\alpha),
\end{equation}
where O3H$\beta$ is the line flux ratio [O{\sc iii}]5007/H$\beta$ and N2H$\alpha$ is the line flux ratio [N{\sc ii}]6584/H$\alpha$. This indicator has the main advantages of i) using emission lines which have very similar wavelengths, thus being essentially independent of dust attenuation and ii) having a unique metallicity for each line flux ratio. Figure \ref{MetalDEP} shows the difference between A$_{\rm H\alpha}$ computed with the empirical [O{\sc ii}]/H$\alpha$ and A$_{\rm H\alpha}$ estimated directly from the Balmer decrement, as a function of metallicity. For comparison, the Calzetti law prediction is also shown, emphasizing that the mismatch between the latter law and the observational data is mostly due to the the effect of metallicity on the [O{\sc ii}]/H$\alpha$ ratio, which is not taken into account by Calzetti, but is incorporated in the empirical calibration presented in this work. The results suggest that even if galaxies at $z=1.47$ have different metallicities from those in SDSS, no significant offset (within the scatter, $\pm0.3$ mag) is expected when estimating A$_{\rm H\alpha}$ from [O{\sc ii}]/H$\alpha$ for a wide range of metallicities.

For the remaining of the analysis, A$_{\rm H\alpha}$ is computed as in equation 12, both for the $z=1.47$ and the SDSS samples. It should be noted that the qualitative and quantitative results remain unchanged if Balmer decrements are used instead to estimate A$_{\rm H\alpha}$ for the SDSS galaxies, and that qualitative results also remain unchanged if the Calzetti law/best linear fit is used instead of the polynomial fit. The sample of H$\alpha$ emitters at $z=0.1$ presents A$_{\rm H\alpha}=0.97\pm0.42$, while the sample of H$\alpha$ emitters at $z=1.47$ presents A$_{\rm H\alpha}=1.0\pm0.6$.

%
%
\begin{figure}
\centering
\includegraphics[width=8.2cm]{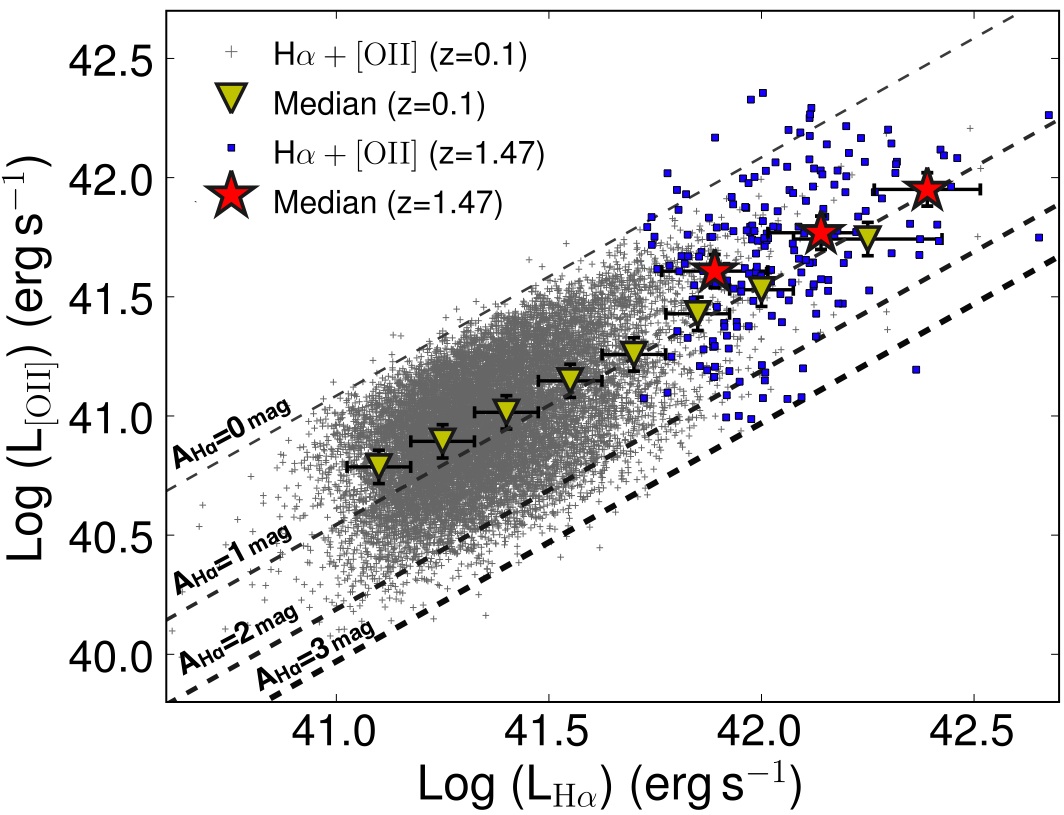}
\caption[Relation between observed luminosities at both redshifts]{A
  comparison between observed [O{\sc ii}] and H$\alpha$ luminosities
  (not corrected for dust-extinction) for samples at $z\sim0.1$ and
  $z\sim1.5$. The correlation between observed luminosities seem to
  evolve very little in the last $\sim9$\,Gyrs, revealing a typical 1
  mag of extinction at H$\alpha$ for the probed
  luminosities. Lines of constant extinction are also shown -- note that these are not evenly spaced due to the use of the non-linear relation (Equation 12). \label{OII_HA_Lumi_correl_SDSS_and_HiZELS}}
\end{figure}

\subsection{[O{\sc ii}]-H$\alpha$ luminosity correlation at $z=1.47$
  and $z \sim 0.1$}  \label{OII_Ha_corr_z147}

Figure \ref{RATIOS} presents the distribution of [O{\sc ii}]/H$\alpha$
line ratios for the $z=1.47$ sample. While it reveals a relatively
wide range within the sample, ($\approx0.08-1.2$), it also shows that
down to the H$\alpha$ flux limit, the line ratio distribution peaks at
$\approx0.5$ (a bit above the median value, $\approx0.45$). In the
SDSS-derived sample, the line ratios show a similar range from $0.1-1$
\citep[c.f.][]{Hopkins_SLOAN} and the median observed line ratio is
found to peak at $\approx0.4$. Thus, even though the H$\alpha$ and [O{\sc
    ii}] luminosities in the $z=1.47$ sample are much higher than
those probed locally, the typical line ratio, if anything, is higher,
indicating lower extinction.

%
%
\begin{figure}
\centering
\includegraphics[width=8.2cm]{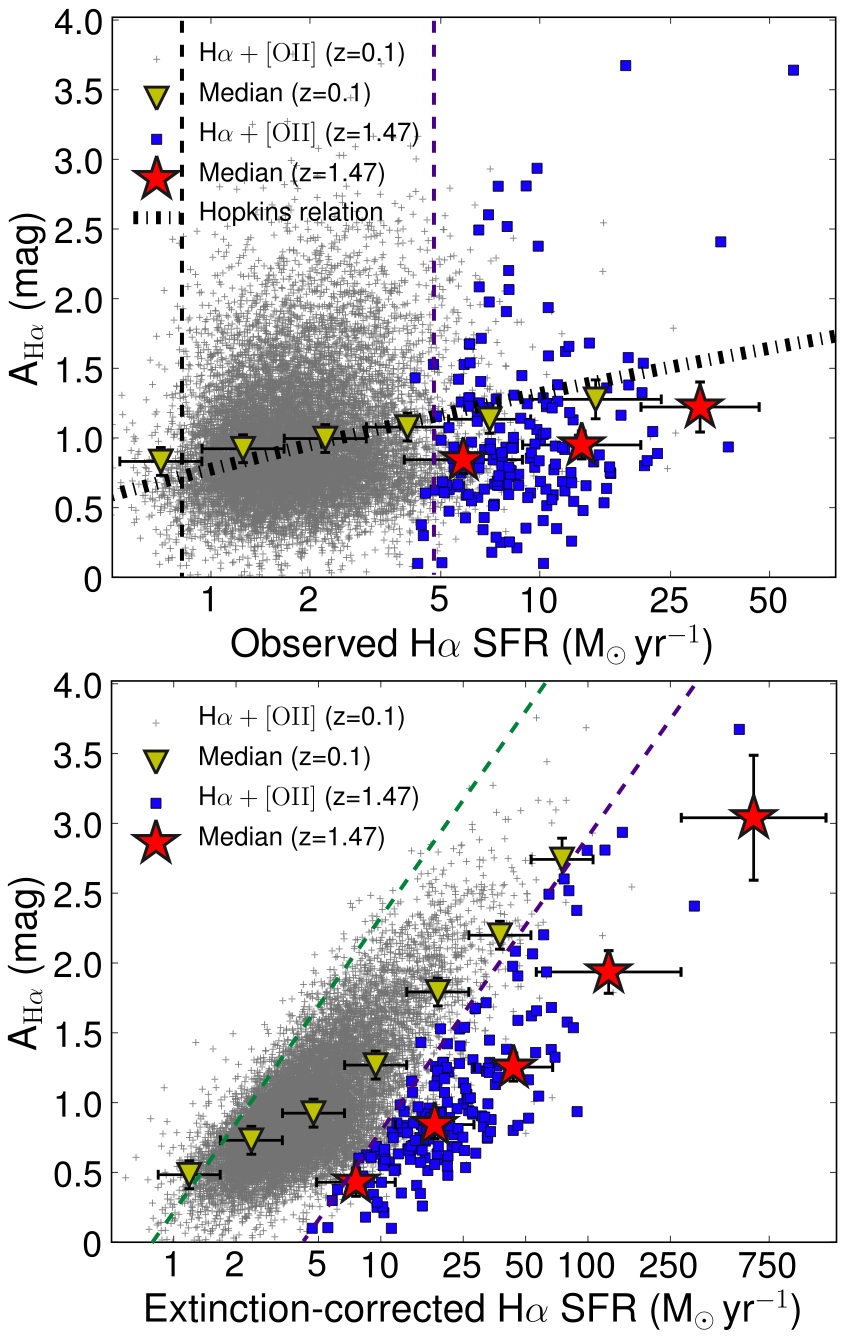}
\caption[line ratios]{The relation between dust extinction and star formation rates (uncorrected and corrected for dust extinction) based on the H$\alpha$ emission lines for both $z=1.47$ and $z=0.1$. The $top$ panel shows the relation between dust extinction at H$\alpha$ and observed H$\alpha$ SFRs. The figure also shows the \cite{Hopkins} relation; this consistently over-predicts the dust extinction correction of the $z=1.47$ sample by $\sim0.5$ mag and, although it agrees better with the $z=0.1$ sample, it over-predicts the slope of the correlation. The $bottom$ panel presents the relation between A$_{\rm H\alpha}$ and dust extinction-corrected SFRs for both redshifts. Note that because both samples are H$\alpha$ selected, there are strong selection biases affecting the H$\alpha$ analysis (dashed lines, resulting from the approximate selection limit in the two samples, and from using A$_{\rm H\alpha}$ to correct SFRs for extinction), but the trends are still recovered even when accounting for such biases. The results show that even though the trends are similar at both $z=0.1$ and $z=1.47$, the normalisation is different, and the trends have evolved between $z\sim1.5$ and $z\sim0$. \label{multipanel_SFR_extinction}}
\end{figure}

This is investigated in more detail in Figure
\ref{OII_HA_Lumi_correl_SDSS_and_HiZELS}, which shows how the
H$\alpha$ and [O{\sc ii}] luminosities (not corrected for extinction)
correlate over the range of luminosities probed by SDSS at $z\sim0.1$
and HiZELS at $z=1.47$. Lines of constant extinction (in H$\alpha$)
are also shown. It is noteworthy that at a given observed H$\alpha$
luminosity, the median [O{\sc ii}] luminosities are slightly
higher (indicating lower median extinction) at the higher redshift.
Both samples show weak trends for increasing extinction with
increasing luminosity. Furthermore, it should be noted that the simple assumption that both samples present a typical constant extinction of 1 magnitude is a relatively good approximation.

\cite{Hopkins} find that there is a correlation between SFR and A$_{\rm H\alpha}$, and so argue that it is possible to estimate (statistical) dust extinction corrections based on observed SFRs, particularly for H$\alpha$ (but also for the [O{\sc ii}]-derived SFRs); their relation has been used to apply statistical corrections to the observed H$\alpha$ luminosities in many studies at low and high redshift. On the other hand, the Hopkins SFR-A$_{\rm H\alpha}$ relation, whilst derived with a relatively small sample in the local Universe, seems to be roughly valid at $z\sim0.84$, as found by \cite{Garn2010a} (which presents a similar analysis to Hopkins et al., i.e. comparing mid-infrared SFRs with H$\alpha$), although it slightly overestimates the amount of dust extinction of the sample. Nevertheless, it is unclear whether a similar result can be found when studying A$_{\rm H\alpha}$ as a function of SFRs at $z=1.47$, or if the result from \cite{Hopkins} can be recovered when using emission-line ratios (e.g. Balmer decrement, or the [O{\sc ii}]/H$\alpha$ calibration) at $z\sim0.1$.

Figure \ref{multipanel_SFR_extinction} presents the results of investigating the dependence of dust-extinction on observed H$\alpha$ SFR for both samples. The results show a relatively weak correlation between A$_{\rm H\alpha}$ and observed H$\alpha$ SFRs at both $z=0.1$ and $z=1.47$; an offset in the median extinction for a given SFR between both epochs is also found.

The figure also shows the \cite{Hopkins} relation between A$_{\rm H\alpha}$ and observed SFRs, with the results showing that it consistently over-predicts the dust extinction correction of the $z=1.47$ sample by $\sim0.5$ mag ($\sim1.6\times$) and, although it agrees better with the $z=0.1$ sample (in the normalisation), it over-predicts the slope of the correlation. Note that A$_{\rm H\alpha}$ at $z=1.47$ are potentially overestimated for extremely metal-poor or metal-rich galaxies, so the offset is robust and, if anything, is underestimated. The SDSS results remain completely unchanged when using the Balmer decrement A$_{\rm H\alpha}$. Therefore the use of the Hopkins et al. relation to correct observed H$\alpha$ SFRs as a function of observed SFRs/luminosity results in a clear overestimation of the dust-extinction correction at $z=1.47$; re-normalising it by $\sim0.5$\,mag is able to solve this.

Nevertheless, there seems to be a correlation between SFR and A$_{\rm H\alpha}$ at $z\sim1.5$, which is more clearly revealed after correcting SFRs for dust extinction, as can be seen in Figure \ref{multipanel_SFR_extinction}. It should however be noted that such relation is in part a result of a bias (as indicated in Figure \ref{multipanel_SFR_extinction}), as for a given flux limit and a (wide) distribution of dust extinction corrections, one easily recovers a relation between extinction and corrected SFRs (as correcting SFR for extinction relies on A$_{\rm H\alpha}$). The relation between extinction and SFR is clear at the highest H$\alpha$ luminosities where such selection biases are negligibly small. The offset between the A$_{\rm H\alpha}$ vs corrected-SFR relations between $z=0.1$ and $z=1.47$ is also clear at these high SFRs, and is still recovered even when a common H$\alpha$ luminosity limit is applied to both samples. These results indicate that although there does appear to be a relationship between dust extinction and SFR, this relation appears to evolve with redshift and should not be used as a reliable way of estimating statistical dust extinction corrections for samples of galaxies at different redshifts.

\subsection{Mass as a dust-extinction indicator}  \label{AHA_dep_MASS}

Recently, \cite{GARNBEST} performed a detailed investigation of the
correlations between dust extinction (A$_{\rm H\alpha}$) and several
galaxy properties (e.g. metallicity, star formation rate, stellar
mass) using large SDSS samples. The authors find that although A$_{\rm H\alpha}$ roughly
correlates with many galaxy properties, stellar mass seems to be the
main predictor of dust extinction in the local Universe
\citep[see also][for a similar analysis]{Gilbank10}. The authors derive a polynomial fit to the
observed trend, which can be used to estimate dust extinction
corrections for galaxies with a given stellar mass in the local
Universe. Nonetheless, so far no study has been conducted in order to test
whether such relation exists at high redshift and whether it evolves
significantly.

\subsubsection{Estimating stellar masses at $z=1.47$}  \label{MASSES_z147}

In order to investigate any potential correlation between dust extinction and stellar mass at $z\sim1.5$, stellar masses are obtained for the entire $z=1.47$ sample, following the methodology fully described in \cite{SOBRAL10B}. Very briefly, the multi-wavelength data available for the $z=1.47$ sources are used to perform a full SED $\chi^2$ fit with a range of models -- normalised to one solar mass -- to each galaxy; the stellar-mass is the factor needed to re-scale the luminosities in all bands from the best model to match the observed data. As in \cite{SOBRAL10B}, the SED templates are generated with the stellar population synthesis package developed by \cite{BC03}, but the models are drawn from \cite{BC07}. SEDs are produced assuming a universal initial mass function (IMF) from \cite{Chabrier03} and an exponentially declining star formation history with the form $e^{-t/\tau}$, with $\tau$ in the range 0.1 Gyrs to 10 Gyrs. The SEDs were generated for a logarithmic grid of 220 ages (from 0.1 Myr to 4.3 Gyr -- the maximum age at $z=1.47$). Dust extinction was applied to the templates using the \cite{Calzetti00} law with $E(B-V)$ in the range 0 to 0.8 (in steps of 0.1). The models are generated with a logarithmic grid of 6 different metallicities, from sub-solar to super-solar metallicity. It is assumed that all H$\alpha$ emitters are at $z=1.47$ and the complete filter profiles are convolved with the generated SEDs for a direct comparison with the observed total fluxes. For all except 5 sources (1 with no $U$ band data, and 4 not detected in any IRAC band), 16 bands are used, spanning from the CFHT $U$ band in the near-ultra-violet to the 4 IRAC bands (M. Cirasuolo et al., in prep.). The appropriate corrections (c.f. Sobral et al. 2011) are applied to obtain total fluxes in each band.


%
%
%
\begin{figure}
\centering
\includegraphics[width=8.2cm]{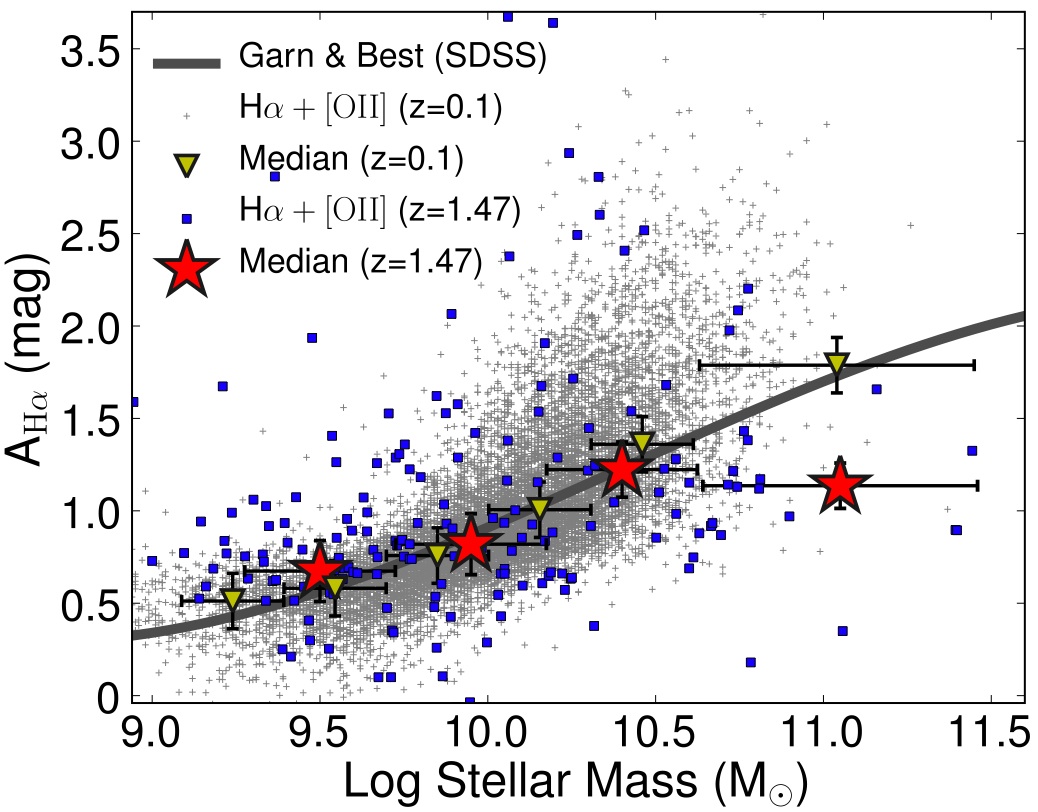}
\caption[Extinction - stellar mass relation]{The relation between dust
  extinction (A$_{\rm H\alpha}$) and stellar mass for the samples at
  $z=0.1$ and $z=1.47$. Individual galaxies from both samples and the
  median are presented. The relation derived by \cite{GARNBEST} is
  also shown, which is found to describe the relation very well for
  $z=0.0-1.5$ at least for low and moderate masses, indicating that stellar mass is likely a fundamental property for predicting dust-extinction corrections. The most massive star-forming galaxies (log Mass$>10.75$) may have different extinction properties at $z=0.1$ and $z=1.5$, although that may well be a result of the small number of such galaxies at $z=1.5$. \label{STELLARMASS}}
\end{figure}

Stellar mass estimates of each individual source are found to be affected by a 1$\sigma$ error (from the multi-dimension $\chi^2$ distribution) of $\sim0.30$ dex, which results from degeneracies between the star formation time-scale $\tau$, age, extinction and, to a smaller extent, metallicity. As the analysis uses a Chabrier IMF, stellar masses are directly comparable with SDSS masses \citep[which are the ones used by][]{GARNBEST}, but note that there is a systematic offset when compared to Salpeter \citep[see][]{SOBRAL10B}. It should also be noted that the $E(B-V)$ from the best fits correlate well with the A$_{\rm H\alpha}$ determined for each individual galaxy.

\subsubsection{Mass-Extinction relation}  \label{MASSES_results}

The sample of $z=1.47$ emitters presents a median stellar
mass of 10$^{9.9}$\,M$_{\odot}$. Figure \ref{STELLARMASS} shows the observed relation between A$_{\rm
  H\alpha}$ and stellar mass, for both $z=0.1$ and $z=1.47$ samples,
together with the \cite{GARNBEST} relation. The results reveal that
not only is there a correlation between stellar mass and dust
extinction at $z=1.47$, just like the one at $z\sim0$, but, even
more importantly, that the \cite{GARNBEST} relation seems to be valid at least up
to $z\sim1.5$ as a dust extinction estimator for most masses. As shown above, this is in contrast with SFR-dependent extinction corrections, which must at least be re-normalised when being applied to $z\sim1.5$ or $z\sim0$, and provides an important insight into what is important in determining the dust properties of galaxies. Table \ref{Extinction_fits} presents the \cite{GARNBEST} relation for predicting A$_{\rm H\alpha}$ as a function of stellar mass, which can be applied, at least within the studied range of masses, to derive statistical corrections for samples of galaxies up to $z\sim1.5$.

Although the relation between stellar mass and A$_{\rm H\alpha}$ seems to hold across redshift for most masses probed, there is tentative evidence of an offset at the highest masses, in the sense that the most massive $z=1.47$ star-forming galaxies appear to be affected by significantly less dust extinction that those with comparable masses at $z=0.1$. The sample at high masses remains relatively small, however, and also since these massive star-forming galaxies are all of very low H$\alpha$ EWs (see Sobral et al. 2011), there is the possibility that selection effects are driving this result. Until this can be investigated with an improved sample, however, caution should be taken in applying the mass-extinction relation at high redshifts for masses above $\sim10^{10.75}$\,M$_{\odot}$.

\subsection{Predicting A$_{\rm H\alpha}$ dust-extinction with colours}  \label{ratio_dep_color}

Stellar mass seems to be a relatively reliable (and important) non-evolving dust-extinction predictor which can therefore be applied both in the local Universe and at higher redshift ($z\sim1.5$), at least within a range a masses. However, computing reliable stellar masses requires obtaining a significant number of observations at different wavelengths, relies on the validity of the stellar population models used to compute mass to light ratios, and it is, in general, a difficult quantity to estimate, particularly at high redshift. There is therefore the need to investigate more direct observables as a way to predict the median dust extinction of a population of galaxies that could be applied out to $z\sim1.5$. Rest-frame optical colours are expected to correlate with dust extinction, and are therefore investigated and calibrated as dust-extinction tracers. All colours presented in this section are given in the AB system and are rest-frame colours computed using the SDSS filters.

%
%
%
\begin{figure}
\centering
\includegraphics[width=8.2cm]{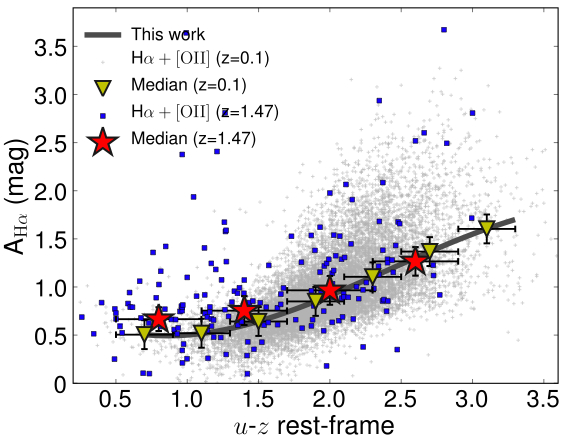}
\caption[The relation between A$_{\rm H\alpha}$ and colour]{The relation between dust extinction and rest-frame $u-z$ colour for the samples at $z=0.1$ (SDSS) and $z=1.47$. The data are used to derive an empirical relation which is valid for H$\alpha$ emitters both at $z=0.1$ and $z=1.47$ and predicts median dust extinction corrections (A$_{\rm H\alpha}$) based on $u-z$ colours. Table \ref{Extinction_fits} presents the best-fit polynomial relations based on other rest-frame colours which also correlate well with A$_{\rm H\alpha}$. \label{OII_HA_extinction_vs_colour}}
\end{figure}

Figure \ref{OII_HA_extinction_vs_colour} presents A$_{\rm H\alpha}$ as a function of rest-frame ($u-z$) for both $z=0.1$ and $z=1.47$. For the $z=1.47$, the observed ($z'-K$) colours\footnote{$z'$-band data from Suprime-cam/Subaru; $K$ from WFCAM/UKIRT.} can be used, as the redshifted $z'$ and $K$ filter profiles broadly match $u$ and $z$ SDSS filters and a simple statistical correction of $+0.15$\footnote{This statistical correction was computed by using the range of best-fit SED models to the z=1.47 sources and measuring the ($z'-K$) colours with the $z'$ and $K$ filters on Subaru and UKIRT, respectively, following by measuring the ($u-z$) rest-frame colours of the same sources, using the $u$ and $z$ SDSS filters.} on the ($z-K$) colour (which accounts for a small K correction and the differences in the filter profiles) is able to match the colours very well.

The results presented in Figure \ref{OII_HA_extinction_vs_colour} show that there is a significant correlation between galaxy colour and dust-extinction and suggest that, despite galaxies in the sample at $z=1.47$ being bluer (on average), a single relation seems to hold across epochs (at least out to $z\sim1.5$). Indeed, a simple polynomial fit to the median extinction for galaxies with a given rest-frame $u-z$ colour ($\approx$ observed $(z'-K)+0.15$ colour at $z\sim1.5$) is valid at both $z=0.1$ and $z=1.47$, and is given by:

\begin{equation}
 {\rm A_{\rm H\alpha}}=-0.092(u-z)^3+0.671(u-z)^2 -0.952(u-z)+0.875
\end{equation}
Relations between A$_{\rm H\alpha}$ and various other optical rest-frame colours are also investigated; these can be a valuable tool to estimate dust-extinction of galaxy populations at different redshfits where only a simple colour is available. Table~\ref{Extinction_fits} presents the best fits to the data that are valid at least up to $z\sim1.5$, together with the limits within the relations are valid. The scatter is also quantified for each fit (see Table~\ref{Extinction_fits}), revealing that the relations provide good fits of comparable quality to the mass-extinction relation. 

%
%
\begin{table}
 \centering
  \caption{Predicting median dust-extinction (A$_{\rm H\alpha}$, mag) corrections using rest-frame colours and mass for
    galaxies at $z\sim0-1.5$. Best-fit relations are based on 3rd order polynomials: $y=Ax^3+Bx^2+Cx+D$. Lower and upper limits of validity of the fits are also presented. Note that the photometry is in the AB system and using SDSS filters and that $Mass$ is given in units of $\log$(M$_{\odot}/10^{10}$). The scatter of the data relative to the best fit is also given.}
  \begin{tabular}{@{}ccccccc@{}}
  \hline
   Property & A & B & C & D & Validity & Scatter \\
 \hline
   \noalign{\smallskip}
 $Mass$ & $-$0.09 & 0.11 & 0.77 & 0.91 & [$-0.4$,0.3] & 0.33 \\
 \bf ($u-g$) & $-$1.31 & 4.59 & $-$4.15 & 1.68 & [0.4,1.5] & 0.35 \\
 \bf ($g-r$) & $-$2.63 & 5.00 & $-$0.78 & 0.51 & [0.0,0.8] & 0.28  \\
 \bf ($r-i$) & $-$27.29 & 29.83 & $-$7.64 & 1.10 & [0.1,0.55] & 0.35 \\
  \bf ($u-z$) & $-$0.092 & 0.671 & $-$0.952& 0.875 & [0.5,3.2] & 0.30 \\
 \hline
\end{tabular}
\label{Extinction_fits}
\end{table}

\subsection{Discussion of emission-line ratios} 

The mean dust extinction properties of the sample of moderately star-forming galaxies at $z\sim1.5$ seem to be very similar to those in the local Universe (a simple 1 mag of extinction at H$\alpha$ for the entire population of $z\sim0.1$ and $z\sim1.5$ galaxies is a relatively good approximation, but with a scatter of 0.3 mag), as a whole, even though modest-SFR galaxies at $z\sim1.5$ seem to be slightly less extinguished. Even more interesting is the
fact that dust extinction presents the same dependence on stellar mass in the last 9\,Gyrs, at least for star-forming galaxies with low and moderate stellar masses. In contrast, while dust extinction correlates with SFRs at both $z=0$ and
$z\sim1.5$, the normalization of the relations clearly evolves, with differences of $\sim0.5$ mag in H$\alpha$ for the same (corrected) SFR.

As extinction-corrected SFRs correlate reasonably well with
stellar mass and dust extinction also correlates with (corrected) SFRs,
it is possible that the relation simply evolves as sSFRs evolve. Physically, the
normalization of the relation could be driven by the gas reservoirs in
galaxies; allowing them to reach much higher SFRs at $z\sim1.5$ than
locally, for a fixed stellar mass. This conclusion is in line with
\cite{GARNBEST} and has important consequences towards understanding
galaxy evolution in the last 9 Gyrs and how little dust properties seem to have changed in galaxies with modest SFRs.

\section{Conclusions}  \label{conclusions}

This paper presented the results from the first panoramic matched
H$\alpha$+[O{\sc ii}] dual narrow-band survey at $z\sim1.5$. This is a very
effective way of assembling large and robust samples of
H$\alpha$ and [O{\sc ii}] emitters at $z=1.47$. It provides a large, robust sample of H$\alpha$ emitters at $z=1.47$,
together with a large sample ($\sim1400$) of [O{\sc ii}] emitters at the
same redshift. The survey has allowed for the first statistical direct comparison of
H$\alpha$ and [O{\sc ii}] emitters at $z\sim1.5$ and a direct
comparison with an equivalent sample in the local
Universe to look for evolution. The main results are:

\begin{itemize}

\item The well-defined samples of emitters were used to compute the
  H$\alpha$ and [O{\sc ii}] luminosity functions at the same
  redshift. For the H$\alpha$ luminosity function at $z=1.47$, the best-fit Schechter function parameters are: $\log L^*_{\rm H\alpha}=42.5\pm0.2$\,erg\,s$^{-1}$, $\log\phi^*_{\rm H\alpha}=-2.4\pm0.3 $\,Mpc$^{-3}$ and $\alpha_{\rm H\alpha}=-1.6\pm0.4$, while for the [O{\sc ii}] luminosity function at the same redshift the best-fit parameters are: $\log L^*_{\rm [O{\sc II}]}=41.71\pm0.09$\,erg\,s$^{-1}$, $\log\phi^*_{\rm [O{\sc II}]}=-2.01\pm0.10$\,Mpc$^{-3}$ and $\alpha_{\rm [O{\sc II}]}=-0.9\pm0.2$.

\item Both H$\alpha$ and [O{\sc ii}] luminosity functions show a strong and consistent evolution in $\phi^*$ and $L^*$ from $z\sim0$ to $z\sim1$ and a continued $L^*$ evolution to $z\sim1.5$ and beyond. By combining the results with other HiZELS measurements and other estimates from the literature, our understanding of the star-formation history of the Universe is improved. Using a single well-calibrated indicator, the star-formation rate density is shown to rapidly increase out to $z\sim1$, and to probably continue rising (although much more weakly) out to $z\sim2$, due to the steep faint-end now measured at $z=2.2$. At $z=1.47$, the H$\alpha$ analysis yields $\rho_{\rm SFR}=0.16\pm0.05$\,M$_{\odot}$\,yr$^{-1}$ Mpc$^{-3}$, while the [O{\sc ii}]  analysis yields $\rho_{\rm SFR}=0.17\pm0.04$\,M$_{\odot}$\,yr$^{-1}$ Mpc$^{-3}$, in excellent agreement.

\item By using SDSS, the [O{\sc ii}]/H$\alpha$ line fraction is
  calibrated as a dust-extinction probe against the Balmer decrement. The relation is shown to be
  accurate within $\approx0.3$ dex.

\item H$\alpha$ and [O{\sc ii}] luminosities correlate well at
  $z=1.47$, similarly to those at $z\sim0.1$, but the sources at higher redshifts appear less dust extinguished for a given observed H$\alpha$ luminosity. A relatively weak correlation between observed SFR and dust extinction is found for both $z=0.1$ and $z=1.47$, but with a different normalisation. It is also shown that the Hopkins relation consistently over-predicts dust extinction corrections for $z\sim1.5$ by $\sim0.5$\,mag in H$\alpha$.
  
  \item Stellar mass is shown to be a good dust-extinction
  predictor, at least for low and moderate mass galaxies, with the relation between dust extinction and mass being the same in the last 9 Gyrs for such star-forming galaxies. The relation between mass and dust
  extinction from Garn \& Best (2010) is shown to be fully valid with
  no evolution at $z=1.47$.
  
  \item  Optical or UV colours are shown to be a simple observable extinction predictor which can be applied for $z\sim0-1.5$ star-forming galaxies; the best-fit relations based on several colours are derived and presented.

\end{itemize}

The results presented in this paper contribute to
our understanding of the nature and evolution of star-forming galaxies at the
likely peak of the star-formation history of the Universe,
particularly by showing an effective and clean way of selecting large
samples of these galaxies and by investigating, for the first time, the evolution using
both the H$\alpha$ and [O{\sc ii}] emission lines at $z\sim1.5$ at the same time. The results show a very good agreement between the H$\alpha$ and [O{\sc ii}] view. Moreover the results reveal that the typical necessary extinction corrections for the probed H$\alpha$ luminosities at $z=1.5$ is
A$_{\rm H\alpha}\approx1.0$ (but with a scatter of $\approx0.3$\,mag); this is what has been found for a
range of luminosities in the local Universe, therefore revealing no
significant evolution for moderate SFR galaxies (if anything, sources at higher redshift are less extinguished). Extinction corrections blindly applied as a function of H$\alpha$ luminosity (e.g. using the local Hopkins relation)
over-predict the dust extinction correction for the $z=1.47$ sample and
would lead to significant biases.

\section*{Acknowledgments}

The authors thank the reviewer for useful comments and suggestions which improved the quality and clarity of the paper. DS acknowledges the Funda{\c c}{\~ao para a Ci{\^e}ncia e Tecnologia
  (FCT) for a doctoral fellowship. PNB acknowledges support from the
  Leverhulme Trust. Y. M. and I. R. S. thank the U.K. Science and Technology
  Facility Council (STFC). J.E.G. acknowledges the Natural Science
  and Engineering Research Council (NSERC) of Canada and STFC. The authors wish
  to highlight the crucial role and unique capabilities of UKIRT and
  the JAC staff in delivering the extremely high-quality data which
  allowed this study to be conducted, and the synergies that can be
  explored between UKIRT, Subaru and other telescopes. The authors
  would also like to thank Masami Ouchi, Tomo Goto, Masao Hayashi, Tadayuki Kodama, Richard Ellis, Andy Lawrence,
  Sebastien Foucaud, Len Cowie, Lisa Kewley, Peder Norberg, Chun Ly and Ester Hu for
  helpful comments and discussions. The authors acknowledge both the
  UKIDSS UDS and the Subaru-SXDS teams for their tremendous effort
  towards assembling the high-quality and unique
  multi-wavelength data-sets which are essential for this paper.

\bibliographystyle{mn2e.bst}
\bibliography{bibliography.bib}

\appendix

\section[]{Completeness Corrections} \label{rComple}

Section \ref{complet} outlines
the procedure which has been followed in order to address such issues
and further details are given in S09. Here details are given on the
effect of the changes in methodology since S09, by addressing how completeness
estimations change with different input samples for the
colour-selection analysis.

%
%
\begin{figure}
\centering
\includegraphics[width=8.2cm]{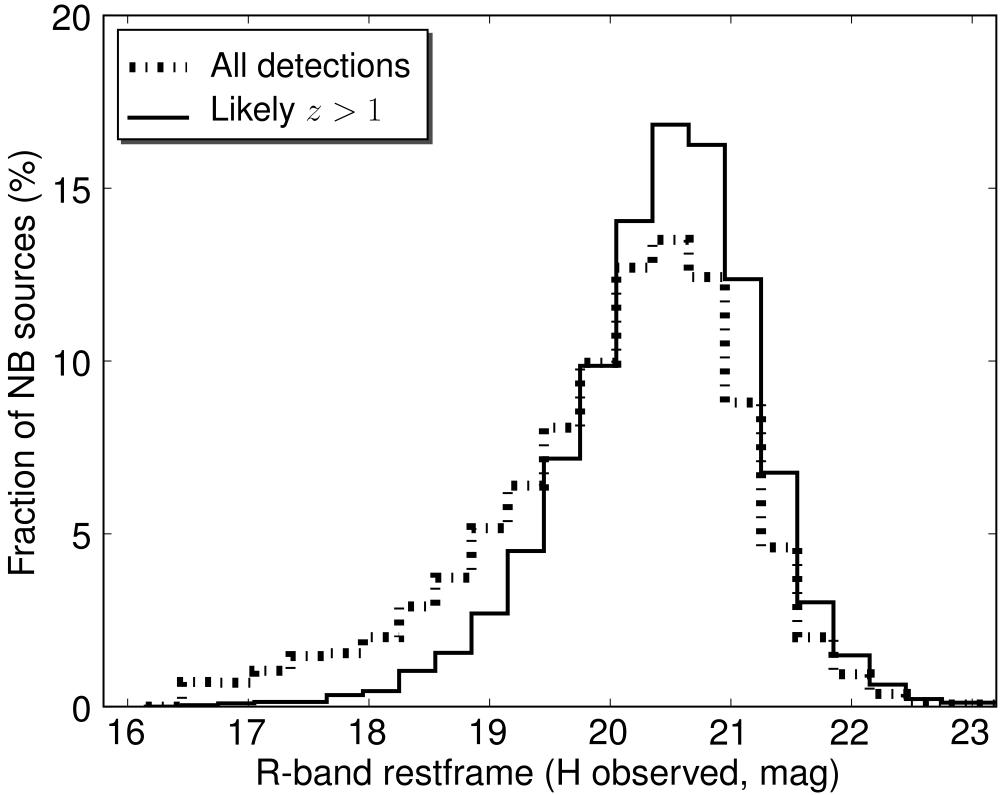}
\caption[Excess Sources in COSMOS]{Observed $H$ band magnitude (R
  rest-frame at $z=1.47$) distributions for the two samples used in
  the completeness simulations. Note the difference between the
  samples at bright magnitudes: while the sample of likely $z>1$
  galaxies presents a strong peak at faint magnitudes and a relatively sharp
  decline for brighter magnitudes, the sample containing all
  detections presents a much broader distribution, including a much
  shallower decay at bright magnitudes.  \label{incompleteness_dist}}
\end{figure}

In S09, the entire range of sources detected in the narrow-band imaging
was used to add line fluxes and then study the recovery rate. This can,
nonetheless, be improved (as detailed in Section \ref{complet}),
to compute completeness corrections specifically for $z\sim1.5$
sources. Here, that is done by i) excluding all stars (since they are
not real $z\sim1.5$ galaxies and they do not have their properties)
and ii) excluding sources with colours which clearly place them
clearly at $z<1$. By taking this approach, the observed H band
magnitude distribution is quite different from that of all detections,
as can be seen in Figure \ref{incompleteness_dist}. Indeed, by using
the entire population of detections, the number of bright sources with low
equivalent widths is overestimated, and thus an overall lower
completeness fraction is derived. Figure
\ref{incompleteness_diff_meth} compares the completeness fraction as a
function of input line flux for simulations using all sources and for
those using only $z>1$ galaxies (rejecting potential stars and using the
$B$$R$$i$$K$ colour-colour selection).

%
%
\begin{figure}
\centering
\includegraphics[width=8.2cm]{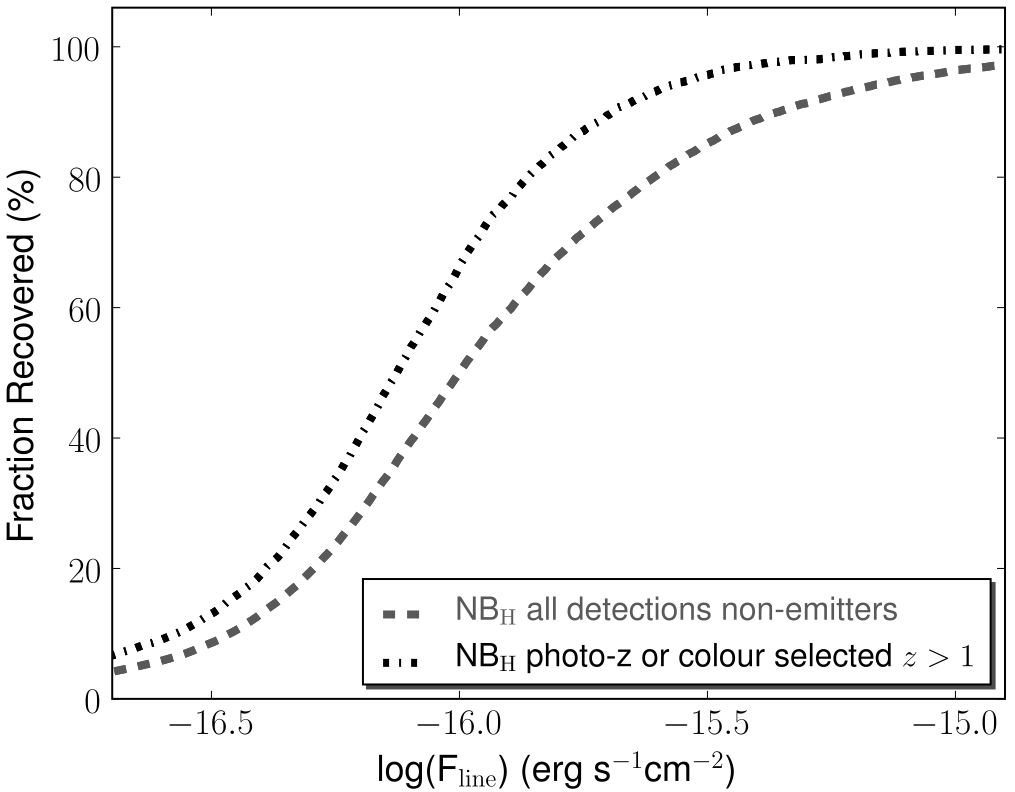}
\caption[Excess Sources in COSMOS]{A study of the completeness
  fraction (defined as the fraction of sources with a given flux
  recovered by the selection against the actual number of source with
  that flux) using two different input samples (all detections which
  are non-emitters and only those consistent with being $z>1$ sources). \label{incompleteness_diff_meth}}
\end{figure}

\bsp

\label{lastpage}

\end{document}